\documentclass[a4paper,10pt]{article}
\usepackage{booktabs, float, jheppub}
\usepackage[dvipsnames,svgnames,table,x11names]{xcolor}
\usepackage[capitalize]{cleveref}
\crefname{figure}{Figure}{Figures}
\usepackage[normalem]{ulem}

\newcommand{\xo}{X_1}
\newcommand{\xt}{X_2}
\newcommand{\met}{E_T^{\mathrm{miss}}}  
\newcommand{\mxo}{m_{X_1}}  
\newcommand{\mxt}{m_{X_2}}  
\newcommand{\U}{\mathrm{U}}
\newcommand{\GL}{\mathrm{GL}}

\usepackage{tikz}
\usetikzlibrary{arrows.meta, decorations.pathreplacing, decorations.markings, patterns, calc}
\usetikzlibrary{decorations.pathmorphing}
\definecolor{cobalt}{rgb}{0.0, 0.28, 0.67}
\usetikzlibrary{decorations.markings}
\usetikzlibrary{arrows.meta}

\tikzset{
pos=.8,
scalar/.style={decorate, dashed, draw=cobalt},
photon/.style={decorate, thick,decoration={snake,segment length=8,post length=1mm,pre length = 1.5mm}},
boson/.style={decorate, thick,draw=cobalt,decoration={snake,segment length=5,post length=1mm,pre length = 1mm}},
particle/.style={draw=black, postaction={decorate}},
fermion/.style={draw=black, postaction={decorate}, decoration={markings,mark=at position .5 with {\arrow[draw=black,scale=1.5,>=stealth]{>}}}},
antifermion/.style={draw=black, postaction={decorate},decoration={markings,mark=at position .5 with {\arrow[draw=black,scale=1.5,>=stealth]{<}}}},
gluon/.style={decorate, draw=black,decoration={coil,amplitude=4pt, segment length=5pt}},
gluon_sm/.style={decorate, draw=black,decoration={coil,amplitude=4pt, segment length=5pt}},
BSM/.style={draw=cobalt, postaction={decorate}, thick,decoration={segment length=8}},
SM/.style={draw=black, postaction={decorate}, thick,decoration={segment length=5}},
}

\newenvironment{diagramBI}{
    \begin{center}
    \begin{tikzpicture}[scale=1.8]
}{
    \end{tikzpicture}
    \end{center}
}

\newenvironment{diagramBFO}{
    \begin{tikzpicture}[scale=1.6]
}{
    \end{tikzpicture}
}

\newcommand\QuarkGluonToX{
\begin{diagramBI}
    \draw[fermion] (0,2.5) -- (1,2.5) node[at start, left] {$q$};
    \draw[gluon] (0,1.5) -- (1,1.5) node[at start, left] {$g$};
    \draw[fermion] (1,2.5) -- (1,1.5);
    \draw[fermion] (1,1.5) -- (2,1.5) node[at end, right] {$j$};
    \draw[photon] (1,2.5) -- (2,2.5) node[at end, anchor=south east, xshift=-0.2cm, yshift=0.0cm] {$\gamma^{\star}/Z^*$};
    \draw[photon] (2,2.5) -- (3,3.0) node[midway, yshift=0.35cm] {$X_{2}$};
    \draw[photon] (2,2.5) -- (3,2.0) node[at end, right] {$X_{1}$};
    \draw[photon] (3.0,3.0) -- (3.6, 3.6) node[at end, right] {$X_{1}$};
    \draw[photon] (3.0,3.0) -- (3.6,2.5) node[at end, right] {$\gamma$};
    
    \filldraw[cobalt] (3.,3) circle (1.5pt);  % Vertex 
    \filldraw[cobalt] (2,2.5) circle (1.5pt);  % Vertex X2 decay   
\end{diagramBI}
}

\newcommand\dmdmToZZ{
\begin{diagramBFO}
    \draw[photon] (0,2.5) -- (1,2.5) node[at start, left] {$X_{1}$};
    \draw[photon] (0,1.5) -- (1,1.5) node[at start, left] {$X_{1}$};
    \draw[photon] (1,2.5) -- (1,1.5) node[xshift=0.4cm, yshift=0.85cm] {$X_{2}$};
    \draw[photon] (1,1.5) -- (2,1.5) node[at end, right] {$Z$};
    \draw[photon] (1,2.5) -- (2,2.5) node[at end, right] {$Z$};   
    \filldraw[cobalt] (1.,2.5) circle (1.5pt);  % Vertex 
    \filldraw[cobalt] (1,1.5) circle (1.5pt);  % Vertex X2 decay   
\end{diagramBFO}
}

\newcommand\dmdmToAA{
\begin{diagramBFO}
    \draw[photon] (0,2.5) -- (1,2.5) node[at start, left] {$X_{1}$};
    \draw[photon] (0,1.5) -- (1,1.5) node[at start, left] {$X_{1}$};
    \draw[photon] (1,2.5) -- (1,1.5) node[xshift=0.4cm, yshift=0.85cm] {$X_{2}$};
    \draw[photon] (1,1.5) -- (2,1.5) node[at end, right] {$\gamma$};
    \draw[photon] (1,2.5) -- (2,2.5) node[at end, right] {$\gamma$};   
    \filldraw[cobalt] (1.,2.5) circle (1.5pt);  % Vertex 
    \filldraw[cobalt] (1,1.5) circle (1.5pt);  % Vertex X2 decay   
\end{diagramBFO}
}

\newcommand\dmdmToZA{
\begin{diagramBFO}
    \draw[photon] (0,2.5) -- (1,2.5) node[at start, left] {$X_{1}$};
    \draw[photon] (0,1.5) -- (1,1.5) node[at start, left] {$X_{1}$};
    \draw[photon] (1,2.5) -- (1,1.5) node[xshift=0.4cm, yshift=0.85cm] {$X_{2}$};
    \draw[photon] (1,1.5) -- (2,1.5) node[at end, right] {$Z,\, \gamma$};
    \draw[photon] (1,2.5) -- (2,2.5) node[at end, right] {$\gamma,\, Z$};   
    \filldraw[cobalt] (1.,2.5) circle (1.5pt);  % Vertex 
    \filldraw[cobalt] (1,1.5) circle (1.5pt);  % Vertex X2 decay   
\end{diagramBFO}
}

\title{Photons, jets and missing momentum from a two-vector dark sector}
\author[a]{Yara do Amaral Coutinho,}
\author[b]{\! Benjamin Fuks,}
\author[b]{\! Mark~D.~Goodsell,} 
\author[c]{\! Bertrand Laforge,} 
\author[c]{\! Jos\'e~Ocariz,} 
\author[d,e,f]{\! Farinaldo~S.~Queiroz}
\author[b,d]{\! and \! Yoxara Villamizar}

\affiliation[a]{Instituto de F\'isica - Universidade Federal do Rio de Janeiro
Av.~Athos da Silveira Ramos 149, Rio de Janeiro - RJ, 21941-972, Brazil}
\affiliation[b]{Laboratoire de Physique Théorique et Hautes Énergies (LPTHE), UMR 7589, Sorbonne Université et CNRS, 4 place Jussieu, 75252 Paris Cedex 05, France}
\affiliation[c]{Laboratoire de Physique Nucléaire et de Hautes \'Energies, UMR 7585 - CNRS IN2P3 - Sorbonne Université - Université Paris Cité, 4 place Jussieu, 75252 Paris Cedex 05, France}
\affiliation[d]{International Institute of Physics, Universidade Federal do Rio Grande do Norte,
Campus Universitario, Lagoa Nova, Natal-RN 59078-970, Brazil}
\affiliation[e]{Millennium Institute for Subatomic Physics at the High-Energy Frontier (SAPHIR) of ANID, Fern\'andez Concha 700, Santiago, Chile}
\affiliation[f]{Departamento de F\'isica, Facultad de Ciencias, Universidad de La Serena,
Avenida Cisternas 1200, La Serena, Chile}

\emailAdd{yara.amaral.coutinho@cern.ch}
\emailAdd{fuks@lpthe.jussieu.fr}
\emailAdd{goodsell@lpthe.jussieu.fr}
\emailAdd{bertrand.laforge@cern.ch}
\emailAdd{ocariz@in2p3.fr}
\emailAdd{farinaldo.queiroz@.ufrn.br}
\emailAdd{yoxara@lpthe.jussieu.fr}

\abstract{
  We investigate the LHC phenomenology of a vector dark-sector effective theory containing two neutral massive vector states, both odd under a dark-parity symmetry. The lightest state is stable and provides a dark-matter candidate, while the leading interactions with the Standard Model arise from dimension-six operators involving the hypercharge field strength. In the prompt-decay regime considered in this work, the heavier state can decay radiatively, leading to a $\gamma+\mathrm{jets}+\met$ signature when the two dark vectors are produced in association with QCD radiation. We study this topology at the LHC through a cut-based analysis, comparing an inclusive missing-transverse-momentum selection with a three-bin strategy that retains coarse shape information. The binned analysis is found to substantially improve the expected reach and probes regions of the parameter space compatible with the observed relic abundance in the standard freeze-out scenario. We also discuss the freeze-in interpretation and the limitations associated with the effective field theory description at high masses.
}
\begin{document}
\maketitle
\flushbottom
%############################################
%############################################
%############################################
%\linenumbers

\section{Introduction \label{sec:int}}
Astrophysical and cosmological observations provide compelling evidence for the existence of dark matter (DM), from galactic to cosmological scales~\cite{Bertone:2004pz, Bertone:2010zza, Bertone:2016nfn, Cirelli:2024ssz}. Galactic rotation curves, gravitational lensing, large-scale structure formation and measurements of the cosmic microwave background together support a consistent picture in which about $26\%$ of the energy density of the Universe is carried by non-baryonic non-luminous matter~\cite{Planck:2018vyg}. Yet the microscopic nature of DM remains unknown, and despite extensive experimental efforts, no conclusive non-gravitational signal has been established so far. Direct-detection experiments, indirect searches using cosmic-ray, gamma-ray and neutrino observations, and collider searches involving missing transverse momentum, displaced vertices or other long-lived-particle signatures have indeed placed stringent constraints on broad classes of models. Identifying the particle nature of DM therefore remains one of the central goals of particle and astroparticle physics.

A broad class of well-motivated scenarios extends the Standard Model (SM) by a dark sector containing new particles and interactions~\cite{Jaeckel:2010ni, Essig:2013lka, Arcadi:2017kky, Batell:2022xau, Arcadi:2024ukq}. If the dark sector is neutral under the SM gauge symmetry, its interactions with ordinary matter can be organised in terms of portal operators built from gauge-invariant combinations of the SM fields. At the renormalisable level, the minimal portals are the vector portal arising from kinetic mixing between a dark $\U(1)$ gauge boson and the SM hypercharge field~\cite{Holdom:1985ag}, the Higgs portal mediated by a scalar singlet coupled to the SM Higgs doublet~\cite{Silveira:1985rk, McDonald:1993ex, Burgess:2000yq}, and the neutrino portal involving a gauge-singlet fermion coupled to the lepton and Higgs doublets~\cite{Falkowski:2009yz}. Each portal leads to qualitatively distinct phenomenology and motivates dedicated experimental search strategies. Beyond the renormalisable level, or when a symmetry of the dark sector forbids the dimension-four portals, the leading interactions between the SM and the dark states arise from higher-dimensional operators suppressed by powers of a new-physics scale $\Lambda$. In such effective field theory (EFT) frameworks, the operator structure controls the strength and the kinematic properties of the connection between the visible and the dark sectors. In addition, depending on the masses, the couplings, the operator scale and the thermal history, the same frameworks can accommodate qualitatively different cosmological DM production mechanisms, including thermal freeze-out and freeze-in, while leading to distinctive DM signatures at colliders and in astrophysical searches.

We now specialise this EFT perspective to a vector dark sector containing a certain number of dark vectors $X_a$. Unlike the canonical dark-photon scenario in which a single dark Abelian gauge boson communicates with the SM through renormalisable kinetic mixing with the hypercharge field strength, we consider a construction in which this dimension-four interaction is absent. This can be enforced by a discrete dark-parity symmetry under which the dark vectors are odd while all SM fields are even. The same symmetry then stabilises the lightest dark vector, allowing it to play the role of a DM candidate. Following~\cite{Aebischer:2022wnl}, we focus on the minimal field content required for this construction, comprising two SM-singlet massive vector bosons $\xo$ and $\xt$, both odd under the dark parity.

With kinetic mixing absent, the leading interactions between the dark vectors and the SM arise at dimension six through operators involving the hypercharge field strength tensor and two dark-vector field strength tensors. A key structural feature of these triple-field-strength operators is that they are antisymmetric in the dark-flavour indices and therefore vanish identically for a single dark-vector species. This is why at least two distinct dark-vector states are required for these operators to mediate interactions with the SM. Assuming $\mxt>\mxo$, the lighter state $\xo$ is stable, whereas the heavier state $\xt$ can decay into $\xo$ in association with an electroweak gauge boson. After electroweak symmetry breaking, the hypercharge field strength decomposes into a photon and a $Z$-boson component. Consequently, both the radiative decay $\xt\to\xo\gamma$ and, if kinematically allowed, the massive channel $\xt\to\xo Z$ can occur. The radiative decay $\xt\to\xo\gamma$ is the cornerstone of the collider phenomenology studied in this work. At hadron colliders, the same dimension-six operators mediate the production of $\xo\xt$ pairs, with additional jets arising from QCD radiation. The subsequent decay of the heavier state then yields a typical final state containing an energetic photon, possibly one or more initial-state-radiation jets, and a substantial amount of missing transverse energy carried by the invisible dark vectors. This topology therefore provides a clean and distinctive experimental signature benefiting from the excellent photon reconstruction and identification capabilities of the LHC detectors.

Final states containing photons and large missing transverse momentum have been extensively studied at the LHC. Inclusive monophoton ($\gamma+\met$) searches~\cite{ATLAS:2020uiq, CMS:2025cgw}, often interpreted in terms of simplified DM models, large extra dimensions or other weakly coupled extensions of the SM, provide strong constraints on scenarios in which invisible particles are produced in association with hard initial-state radiation. The ATLAS and CMS collaborations have also performed dedicated analyses targeting $\gamma+\text{jets}+\met$ final states, primarily with interpretations in gauge-mediated supersymmetry-breaking scenarios where the photons typically originate from neutralino decays in the cascades of heavier coloured or electroweak states~\cite{CMS:2019agj, ATLAS:2025uij}. Complementary searches have considered final states with one or multiple photons and large missing energy, including categories with additional jets~\cite{CMS:2019oou, ATLAS:2021jbf, ATLAS:2022ckd}. While these analyses share the same reconstructed final-state topology as the signal considered here, the underlying production and decay mechanisms are fundamentally different. In our model, the photon originates from the radiative decay of the heavier dark vector $\xt\to\xo\gamma$ which is induced by dimension-six operators, while the missing transverse momentum is issued from the invisible dark vectors. In addition, since the signal does not involve new coloured states, the accompanying jet activity arises entirely from QCD radiation rather than from the cascade decays of heavy coloured particles. These differences manifest in the relevant kinematic distributions of the signal: in particular, the missing transverse energy spectrum emerging from dimension-six operators is not necessarily optimised by analyses designed for supersymmetric benchmarks. A dedicated sensitivity study, with an event selection tailored to the kinematics of the vector-DM signal, is therefore warranted.

We investigate the LHC sensitivity to this scenario with an analysis targeting the $\gamma+\text{jets}+\met$ final state at a centre-of-mass energy of $\sqrt{s}=13.6~\mathrm{TeV}$, following the object definitions and event-selection strategies used in typical ATLAS searches. The study is based entirely on Monte Carlo simulations of the signal and the relevant SM backgrounds. We design a cut-based event selection, construct exclusive signal regions in missing transverse momentum and estimate the expected exclusion reach using a binned profile-likelihood analysis with the $\mathrm{CL}_s$ method~\cite{Read:2002hq, Cowan:2010js}. Results are presented for two integrated-luminosity benchmarks, $139~\mathrm{fb}^{-1}$ and $3000~\mathrm{fb}^{-1}$. The former is chosen to allow direct comparison with published Run~2 results, while the latter corresponds to the projected high-luminosity LHC (HL-LHC) dataset. Both sets of predictions are evaluated at the LHC Run~3 centre-of-mass energy of $13.6~\mathrm{TeV}$, allowing us to compare the expected reach for a Run-2-sized dataset with that of the high-luminosity LHC within a common collider-energy setup.

The paper is organised as follows. In \cref{sec:mod}, we introduce the effective Lagrangian describing the model, define the dimension-six operators relevant for the LHC phenomenology, and derive analytical expressions for the decay widths of the dark vectors. In \cref{sec:relic}, we discuss the relic-density calculation for both the freeze-out and freeze-in production regimes and identify the cosmologically motivated regions of the parameter space. The Monte Carlo setup, object definitions and event-selection strategy are next presented in \cref{sec:MC_CPheno}, while in \cref{sec:stats} we describe the statistical analysis and derive the expected sensitivity of the LHC to the signal. We summarise our findings and conclude in \cref{sec:conclusions}.

%############################################
%############################################
%############################################

\section{An effective theory for a two-vector dark sector \label{sec:mod}}
We consider a dark sector containing two massive vector fields $\xo$ and $\xt$, both singlets under the SM gauge group. Their field strength tensors are defined conventionally by $X_{a}^{\mu\nu} = \partial^\mu X_a^\nu - \partial^\nu X_a^\mu$, with $a=1,2$. We impose a discrete dark parity under which the dark vectors are odd while all SM fields are even. This symmetry forbids the renormalisable kinetic-mixing operators $B_{\mu\nu}X_a^{\mu\nu}$ between the hypercharge field strength $B_{\mu\nu}$ and each individual dark vector, and stabilises the lightest dark vector state against decay. In the most general quadratic Lagrangian for two dark vectors, an off-diagonal dark-vector kinetic mixing term $X_{1\mu\nu} X_2^{\mu\nu}$ is however permitted by the symmetries of the theory since it is even under the dark parity. Provided the kinetic matrix is non-singular and positive definite, such a term can always be removed by a $\GL(2,\mathbb{R})$ field redefinition of the dark-vector doublet $(X_1, X_2)$, followed by an orthogonal rotation to diagonalise the associated mass matrix. We therefore work directly in the mass-eigenstate basis with canonical kinetic and mass terms. Throughout this work, we subsequently treat the corresponding masses $\mxo$ and $\mxt$ as free parameters of the effective theory,  further assuming $\mxt>\mxo$ so that $\xo$ is the stable DM candidate. The microscopic origin of the masses, for instance through a Stueckelberg mechanism or spontaneous symmetry breaking in a dark Higgs sector, is left unspecified, and we assume that any additional states associated with mass generation are either sufficiently heavy or sufficiently weakly coupled that they do not affect the collider and cosmological observables considered below.

Focusing on the dimension-six interactions between the dark vectors and the SM hypercharge field strength, the effective Lagrangian reads,
\begin{equation}\label{eq:lagrangian}\begin{split}
  \mathcal{L} = &\ \mathcal{L}_{\mathrm{SM}}
    -  \frac{1}{4} X_{1}^{\mu\nu} X_{1\mu\nu}
    + \frac{\mxo^2}{2} X_{1}^{\mu} X_{1\mu}
    - \frac{1}{4} X_{2}^{\mu\nu} X_{2\mu\nu}
    + \frac{\mxt^2}{2} X_{2}^{\mu} X_{2\mu}\\ &\qquad \
    + \frac{c_B}{\Lambda^2}\, B_{\mu}^{\ \nu} X_{1\nu}^{\ \ \alpha} X_{2\alpha}^{\ \ \mu}
    + \frac{\tilde{c}_B}{\Lambda^2}\, \widetilde{B}_{\mu}^{\ \nu} X_{1\nu}^{\ \ \alpha} X_{2\alpha}^{\ \ \mu},
\end{split}\end{equation}
where $B_{\mu\nu}$ is the field-strength tensor of the SM $\U(1)_Y$ hypercharge gauge field, and $\widetilde{B}_{\mu\nu}$ represents its dual. The parameters $c_B$ and $\tilde{c}_B$ are dimensionless Wilson coefficients, while $\Lambda$ denotes the effective scale suppressing the dimension-six operators. After electroweak symmetry breaking, these interactions induce couplings of the dark vectors to both the photon and the $Z$-boson field strengths $A_{\mu\nu}$ and $Z_{\mu\nu}$, since $B_{\mu\nu} = c_W A_{\mu\nu} - s_W Z_{\mu\nu}$ with $c_W$ and $s_W$ standing for the cosine and sine of the electroweak mixing angle. The two dimension-six operators in \cref{eq:lagrangian} have distinct CP properties: the operator involving $B_{\mu\nu}$ is CP-even, whereas the one involving $\widetilde B_{\mu\nu}$ is CP-odd. Both structures can contribute to dark-vector production and to the decay of the heavier state. For inclusive quantities such as the heavier dark vector radiative two-body decay width and several total production or annihilation rates, the two operators contribute only through the combination $c_B^2+\tilde c_B^2$, with no interference term. Differential observables, however, may depend on the Lorentz structure of the operator in a non-trivial way, and the CP-even and CP-odd interactions can in principle lead to different kinematic distributions. Nevertheless, possible interference effects would require CP-sensitive observables or momentum correlations which are not analysed in this work. In our numerical analysis we then adopt the benchmark choice
\begin{equation}\label{eq:benchmark}
  c_B=\tilde c_B=1,
\end{equation}
which treats the CP-even and CP-odd operators on equal footing. With this convention, the overall interaction strength is controlled by the EFT scale $\Lambda$, and the three-dimensional model's parameter space is defined by the two masses $\mxo$ and $\mxt$ and $\Lambda$.

The Lorentz contraction of the triple-field-strength structures in \cref{eq:lagrangian} makes the corresponding operators antisymmetric under the interchange of the dark-vector flavour indices. They therefore vanish identically when both dark vectors belong to the same species. This explains why at least two distinct dark vectors are required for these interactions to be present at all, and therefore why the minimal two-state case considered is the simplest viable setup. Other dark-parity-even dimension-six operators, such as operators involving the SM Higgs doublet $H$ and one or two dark-vector field strengths like $(H^\dagger H)X_{\mu\nu}^aX^{b\mu\nu}$ and their dual analogues, can in principle contribute as well. In the simplified setup considered here, we set these additional operators to zero and retain only the hypercharge-field-strength operators responsible for the photon-mediated collider signature under scrutiny. We also recall that the whole EFT description is reliable only when the characteristic momentum transfer of the considered processes remains below the mass scale of the heavy degrees of freedom that generate the dimension-six operators. This point is particularly relevant for LHC production where the hard scale of the partonic collision can enter the TeV range. In the collider analysis below we use $\Lambda=1~\mathrm{TeV}$ as a benchmark value, together with other larger choices, and the high-energy tails of the signal distributions must therefore be interpreted with care. The reconstructed photon, jet and missing-transverse-momentum spectra studied in \cref{sec:MC_CPheno} illustrate the kinematic regime selected by the analysis, although they do not by themselves constitute a sharp EFT-validity criterion. A fully conservative EFT treatment would indeed require either an explicit truncation of events probing momentum transfers above the cutoff scale, or a UV completion specifying the heavy states integrated out of the EFT. We leave such refinements for future work and interpret the results presented here as a benchmark sensitivity study within the effective description.

The electroweak decomposition of the hypercharge field strength $B_{\mu\nu}$ implies that the heavier dark vector can decay through the two channels  $\xt \to \xo \gamma$ and $\xt \to \xo Z$, with the latter being kinematically accessible only when $\mxt-\mxo>m_Z$. Within the simplified operator basis of \cref{eq:lagrangian} and neglecting off-shell three-body decays mediated by a virtual photon or $Z$ boson, no other decay modes are generated at leading order, so the $\xt$ total width is given by $\Gamma_{\xt} = \Gamma(\xt\to\xo\gamma) + \Gamma(\xt\to\xo Z)$ and the branching fractions are shared exclusively between these two channels. The corresponding two-body partial widths are
\begin{equation}\label{eq:x2decays}\begin{split}
  \Gamma(\xt \to \xo \gamma) = &\ \frac{c_W^2\,\mxt^5}{96\pi\Lambda^4} \left(c_B^2+\tilde c_B^2\right)
(1+\varrho)(1-\varrho)^3, \\[2mm]
  \Gamma(\xt \to \xo Z) = &\ \frac{s_W^2\,\mxt^5}{96\pi\Lambda^4} \lambda^{1/2}(1,\varrho,z) \bigg[ c_B^2 \Big( (1+\varrho+z)\lambda(1,\varrho,z) \Big)\\
   &\hspace{1.7cm} + \tilde c_B^2 \Big( 1-(\varrho+z)-(\varrho^2+z^2) +(\varrho^3+z^3) +(6-\varrho-z)\varrho z \Big) \bigg],
\end{split}\end{equation}
where
\begin{equation}
  \varrho=\frac{\mxo^2}{\mxt^2},
  \qquad
  z=\frac{m_Z^2}{\mxt^2}
  \qquad\text{and}\qquad
  \lambda(a,b,c)=a^2+b^2+c^2-2(ab+ac+bc).
\end{equation}
These expressions show that no $c_B\tilde c_B$ interference term contributes to the spin-summed two-body decay widths. The CP-even and CP-odd operators therefore enter incoherently: in the radiative channel this dependence reduces to the simple combination $c_B^2+\tilde c_B^2$, whereas in the $Z$ channel the two squared contributions are weighted by different kinematic functions.

Several features are worth noting. First, both partial widths scale as $\mxt^5/\Lambda^4$, as expected for a two-body decay induced by a dimension-six operator. The lifetime of $\xt$ is therefore very sensitive to both the heavy-vector mass and the EFT scale, which has direct implications for the validity of the prompt-decay assumption central to the collider analysis performed in this work. We will quantify this condition explicitly when presenting the parameter-space coverage in \cref{sec:MC_CPheno}. Second, the radiative partial width contains the phase-space factor $(1+\varrho)(1-\varrho)^3$, which vanishes as $(1-\varrho)^3$ in the compressed regime with $\mxt\simeq\mxo$. This suppression reflects both the reduced available phase space and the derivative nature of the field-strength interaction. In the $\xt$ rest frame, the photon energy is then given by $E_\gamma^\ast = (\mxt^2-\mxo^2)/(2\mxt) = (1-\varrho) \mxt/2$, so the same mass splitting that controls the $\xt$ width also controls the hardness of the final-state photon although its properties are further shaped by the boost of the parent dark vector. Third, the decay into an on-shell $Z$ boson is governed by the additional kinematic factor $\lambda^{1/2}(1,\varrho,z)$ and is absent when $\mxt-\mxo<m_Z$. In this compressed regime and in the simplified setup considered, $\xt\to\xo\gamma$ is the only two-body decay and its branching ratio is equal to unity once off-shell three-body modes are neglected. Once the $Z$ channel opens, the branching ratio into photons is reduced by the competition between the two modes, weighted by $c_W^2$ versus $s_W^2$ and the respective kinematic factors. 

\begin{figure}
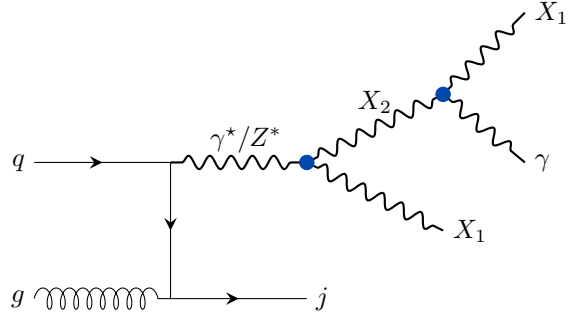

    \centering
    \QuarkGluonToX
    \caption{Representative Feynman diagram for the collider signal considered in this study. An off-shell electroweak gauge boson produces an $\xo\xt$ pair in association with QCD radiation, followed by the radiative decay $\xt\to\xo\gamma$. \label{fig:XtoXgB}}
\end{figure}

The same operators that control the decays of the $\xt$ state also mediate the associated production of the two dark vectors at hadron colliders. After electroweak symmetry breaking, the relevant partonic subprocesses proceed through an off-shell electroweak gauge boson, $q\bar q \to \gamma^\ast/Z^\ast \to \xo\xt$, with additional initial-state QCD radiation generating one or more final-state jets. In this work, we focus on the region of parameter space where the radiative $\xt$ decay is prompt on detector scales. The signal topology therefore comprises a prompt photon, at least one jet and missing transverse momentum carried by the two invisible $\xo$ particles, as illustrated in the representative Feynman diagram of~\cref{fig:XtoXgB}. Scenarios in which $\xt$ is long-lived, leading for instance to displaced photons or other long-lived-particle signatures, are left outside the scope of the present analysis.

%############################################
%############################################
%############################################
\section{Relic abundance of vector dark matter} \label{sec:relic}

The dark parity introduced in \cref{sec:mod} renders the lightest vector state $\xo$ stable, making its relic abundance a key cosmological observable, and we use the value measured by the Planck Collaboration $\Omega_{\rm DM}h^2 = 0.120 \pm 0.001$~\cite{Planck:2018vyg} as a reference for the observed DM abundance. In a standard cosmological history, benchmark scenarios for which $\Omega_{\xo}h^2>\Omega_{\rm DM}h^2$ overproduce dark matter and are therefore disfavoured, whereas points with $\Omega_{\xo}h^2\leq\Omega_{\rm DM}h^2$ remain viable if $\xo$ constitutes all or only a fraction of the observed DM abundance. In the simplified setup considered here, the same dimension-six operators that control the collider signature of the model also govern the production and depletion of the dark-sector population in the early Universe: the relic density is consequently determined by the dark-vector masses $\mxo$ and $\mxt$, the EFT scale $\Lambda$ and the thermal history of the Universe. We consider two limiting DM production mechanisms, the freeze-out and freeze-in ones, which probe qualitatively different regimes of the model. 

\begin{figure}
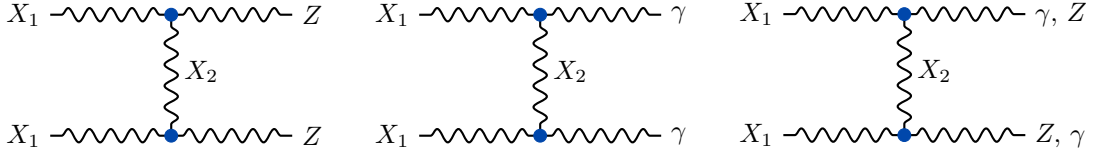

    \centering
    \dmdmToZZ \dmdmToAA \dmdmToZA
    \caption{Representative Feynman diagrams for the annihilation of the stable dark vector $\xo$ into pairs of neutral electroweak gauge bosons, $\xo\xo \to ZZ$ (left), $\gamma\gamma$ (centre) and $Z\gamma$ (right), here mediated by the $t$-channel exchange of the heavier dark vector $\xt$.} \label{fig:dmdmtovv}
\end{figure}

In the freeze-out regime~\cite{Gondolo:1990dk, Griest:1990kh}, the dark sector is assumed to have been in thermal equilibrium with the SM plasma at early times. As the Universe cools down, the interaction rate drops below the Hubble expansion rate and the $\xo$ abundance freezes out at a value controlled by the thermally averaged annihilation cross section at that epoch. In the model discussed in \cref{sec:mod}, the leading annihilation channels into electroweak gauge bosons are $\xo\xo\to\gamma\gamma$, $Z\gamma$ and $ZZ$, proceeding through $t$-channel and $u$-channel exchanges of the heavier state $\xt$ and with representative Feynman diagrams shown in \cref{fig:dmdmtovv}. Since the dimension-six $\xo$-$\xt$-$\gamma/Z$ vertex enters twice in these amplitudes, the corresponding annihilation cross section scales as $\sigma v \propto \Lambda^{-8}$, up to mass-dependent kinematic factors. For fixed dark-vector masses, increasing $\Lambda$ suppresses the annihilation rate and therefore increases the predicted relic density. Conversely, smaller values of $\Lambda$ enhance the annihilation rate and can lead to an underabundant $\xo$ population. If the spectrum is compressed, \textit{i.e.}\ if the mass splitting $\mxt-\mxo$ is small compared to the freeze-out temperature, the thermal abundance of the $\xt$ state at freeze-out is not Boltzmann suppressed relative to that of the DM candidate $\xo$. Co-annihilation processes can then contribute to the effective annihilation rate~\cite{Griest:1990kh, Edsjo:1997bg}. In our case, the relevant co-annihilation channels involve an initial $\xo\xt$ pair annihilating through an off-shell electroweak gauge boson into SM final states, such as fermion pairs, $W^+W^-$ or an associated $Zh$ pair, when kinematically accessible. These processes contain a single insertion of the dimension-six operator, and they may therefore become important in the nearly degenerate regime. After freeze-out, any remaining $\xt$ population eventually decays into $\xo$ through the $\xt\to\xo\gamma$ or $\xt\to\xo Z$ modes, thereby contributing to the final stable $\xo$ abundance.

\begin{figure}
    \centering
    \includegraphics[width=0.5\linewidth]{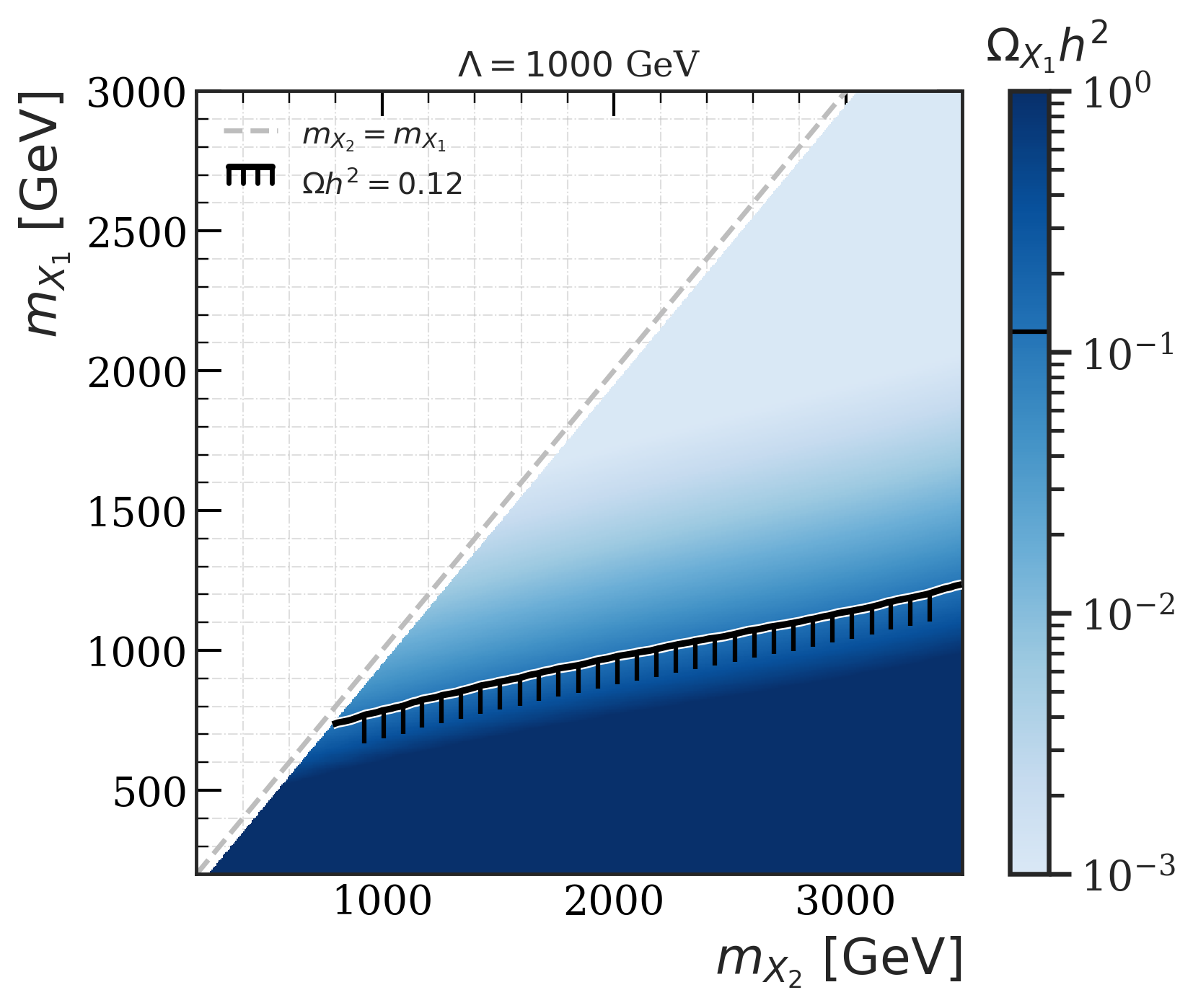}
  \caption{Freeze-out relic abundance in the $(\mxo,\mxt)$ plane for $c_B=\tilde c_B=1$ and $\Lambda=1~\mathrm{TeV}$. The scan is restricted to the region $\mxt>\mxo$, consistently with $\xo$ being the lightest dark-parity-odd state. The black curve indicates $\Omega_{\xo} h^2=0.12$. The region with $\Omega_{\xo} h^2>0.12$ (dark blue) overproduces DM in a standard thermal history and is disfavoured, while the region with $\Omega_{\xo} h^2\leq0.12$ (light blue) is cosmologically allowed if $\xo$ constitutes all or part of the observed DM abundance.}\label{fig:freezeout_relic}
\end{figure}

We evaluate the freeze-out relic abundance numerically with \textsc{MadDM}~\cite{Backovic:2015cra, Ambrogi:2018jqj}, using an implementation of the two-dark-vector effective theory described in \cref{sec:mod} in \textsc{FeynRules}~\cite{Christensen:2009jx, Alloul:2013bka} that is then exported in the UFO format~\cite{Degrande:2011ua, Darme:2023jdn}. The relic-density computation includes the annihilation and co-annihilation channels generated by the model implementation, therefore also including processes involving the $\xt$ state in regions where the spectrum is sufficiently compressed. In the numerical scan we fix the Wilson coefficients to the benchmark values $c_B=\tilde c_B=1$ defined in \cref{eq:benchmark}, set $\Lambda$ to $1~\mathrm{TeV}$ and restrict the scan to the parameter space region in which $\mxt>\mxo$, consistently with the convention that $\xo$ is the lightest dark-parity-odd state and therefore the DM candidate. For each point in the $(\mxo,\mxt)$ plane, we compute the predicted relic density $\Omega_{\xo} h^2$ and compare it with the Planck value $\Omega_{\rm DM}h^2\simeq0.12$~\cite{Planck:2018vyg}.

The result is shown in \cref{fig:freezeout_relic}, where only the region $\mxt>\mxo$ is retained in the analysis. The black curve corresponds to the $\Omega_{\xo} h^2=0.12$ isoline in relic abundance. Correspondingly, points below this curve, shown by the dark blue region, overproduce DM in a standard thermal history and are therefore disfavoured. This overabundant region at small $\mxo$ reflects the strong kinematic suppression of the dimension-six operators at low energies. For $\mxo \ll \Lambda$, the annihilation cross section is indeed too small to deplete the $\xo$ population efficiently in the early Universe. By contrast, points with $\Omega_{\xo} h^2\leq0.12$, shown in light blue, remain cosmologically viable, either because $\xo$ saturates the observed DM density on the contour or because it constitutes only a subcomponent of the dark matter. The structure of the relic-density map follows from the parametric behaviour of the annihilation rate. For fixed $\mxo$, increasing $\mxt$ suppresses the $t$-channel and $u$-channel exchange of the heavier dark vector in $\xo\xo$ annihilation, thereby reducing the annihilation efficiency and increasing the relic abundance. Conversely, increasing $\mxo$ at a fixed $\mxt$ value raises the characteristic energy scale entering the higher-dimensional interaction and can enhance the annihilation rate. Finally, close to the diagonal where $\mxt\simeq\mxo$, co-annihilation effects involving the thermally populated $\xt$ state become relevant, generically enhancing the effective annihilation cross section and slightly reducing the relic density relative to the $\xo\xo$-only result. 

For the benchmark scale $\Lambda=1~\mathrm{TeV}$, a large part of the displayed parameter space is found underabundant. Such points are not excluded by the relic-density measurement constraints, but they would require either an additional DM component or a non-standard cosmological history if the model configuration is to account for all of the DM. In what follows, the freeze-out contour is therefore used as a cosmological reference when comparing with the collider sensitivity, rather than as a requirement that the vector state saturates the observed DM abundance everywhere in the scan. Finally, we note that for other values of $\Lambda$, the same qualitative behaviour is expected. Increasing $\Lambda$ suppresses the annihilation and co-annihilation rates, thereby shifting the observed-abundance contour toward regions with more efficient depletion, namely larger $\mxo$ and/or smaller $\mxt$.

In the freeze-in scenario~\cite{McDonald:2001vt,Hall:2009bx}, the dark sector never reaches thermal equilibrium with the SM plasma. The dimension-six interactions are sufficiently feeble that the dark-sector number densities remain far below their equilibrium values throughout the thermal history, and the $\xo$ abundance is built up gradually through rare scattering processes in the SM bath. Unlike freeze-in through renormalisable interactions, production through higher-dimensional operators is typically UV dominated and can depend sensitively on the reheating temperature~\cite{Elahi:2014fsa}. In the present model, the relevant processes thus depend on the hierarchy between $\mxo$, $\mxt$ and the reheating temperature $T_{\rm rh}$~\cite{Aebischer:2022wnl}. If $\mxo<\mxt<T_{\rm rh}$, both dark vectors are kinematically accessible in the thermal bath. In this regime, associated production processes such as $f\bar f\to\xo\xt$ mediated by a single insertion of the dimension-six operator in \cref{eq:lagrangian} can dominate the freeze-in yield. These one-insertion processes are less suppressed by the EFT scale than the two-insertion processes. Consequently, reproducing the observed relic abundance generally requires a larger suppression scale, making the corresponding collider production rates strongly suppressed. If instead $\mxo<T_{\rm rh}<\mxt$, the heavier state is not efficiently produced from the thermal bath, and the $\xo$ population is generated predominantly through two-insertion processes such as $\gamma\gamma\to\xo\xo$, mediated by virtual $\xt$ exchanges. These processes correspond to the diagrams of \cref{fig:dmdmtovv} read in reverse and have rates scaling as $\Lambda^{-8}$. This stronger suppression leads to smaller cosmologically motivated values of the EFT scale and therefore defines a freeze-in regime more relevant for collider phenomenology. We therefore focus on this hierarchy in what follows. For the benchmark reheating temperatures considered below, $T_{\rm rh}=7$ and $20~\mathrm{GeV}$, this implies that $\mxo$ lies in the GeV range and $\mxt$ in the TeV range, in contrast with the freeze-out case where both dark vectors are typically heavy.

In both regimes, the observed relic abundance fixes a relation among $\mxo$, $\mxt$, $\Lambda$ and $T_{\rm rh}$, rather than selecting a unique value of $\Lambda$ independently of the thermal history. For reheating temperatures below the electroweak scale and in the hierarchy $\mxo<T_{\rm rh}<\mxt$, the analytic estimate of~\cite{Aebischer:2022wnl} gives
\begin{equation}
  \frac{\Lambda}{(c_B^2+\tilde c_B^2)^{1/4}}
    \simeq 700\, T_{\mathrm{rh}} \bigg(\frac{\mxo T_{\mathrm{rh}}^{3}}{10^{3}\, \sqrt{g_{\star}^{3}}~\mathrm{GeV}^{4}}\bigg)^{1/8} \ \bigg(\frac{1~\mathrm{TeV}}{\mxt}\bigg)^{1/2},
\label{eq:freezein_lambda_estimate} \end{equation}
where $g_\star$ denotes the effective number of relativistic degrees of freedom evaluated at $T\sim T_{\rm rh}$. This expression assumes a negligible initial dark-sector abundance and DM production occurring predominantly near the highest temperature of the radiation bath. The strong dependence on $T_{\rm rh}$ reflects the UV-dominated nature of the freeze-in mechanism through higher-dimensional operators. A hotter bath produces more dark vectors, and therefore requires a larger value of the effective scale to reproduce the observed relic abundance.

\begin{figure}
    \centering
	\includegraphics[width=0.5\linewidth]{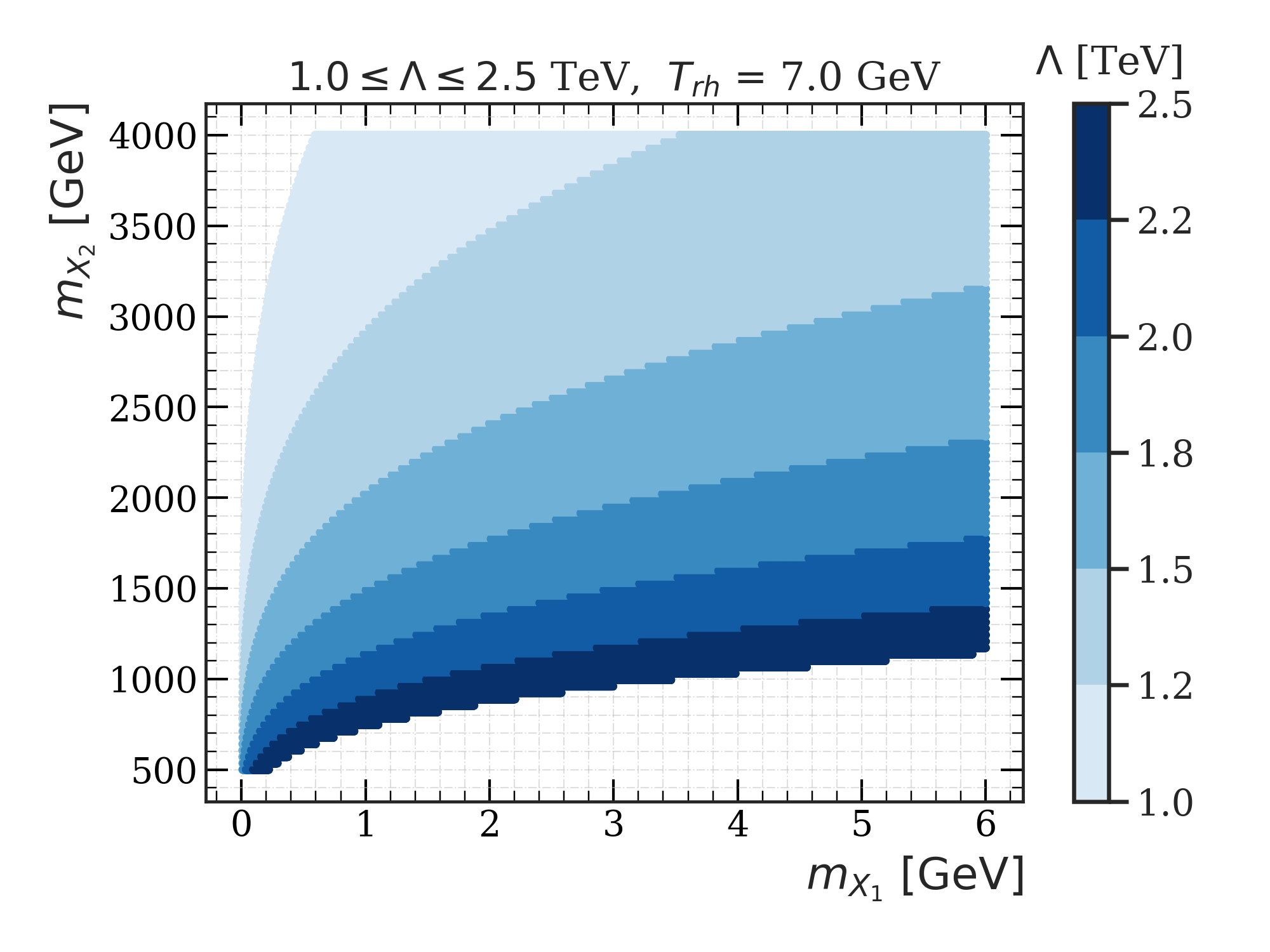}\hfill
    \includegraphics[width=0.5\linewidth]{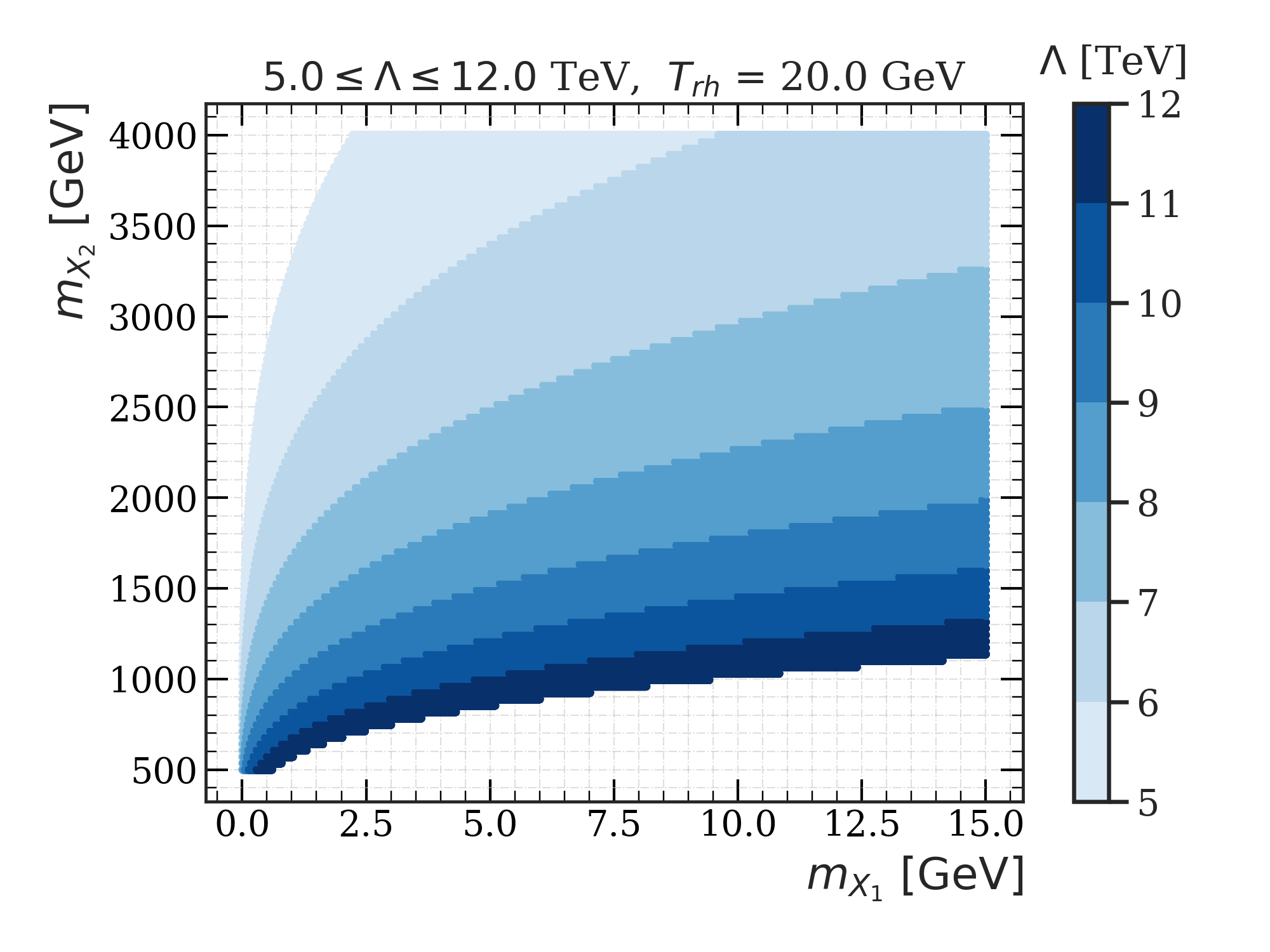}
    \caption{Freeze-in benchmark contours showing the effective scale $\Lambda$ required to reproduce $\Omega_{\xo} h^2\simeq0.12$ in the $(\mxo,\mxt)$ plane, for $T_\mathrm{rh}=7~\mathrm{GeV}$ (left) and $T_\mathrm{rh}=20~\mathrm{GeV}$ (right). The scan covers the hierarchy $\mxo<T_\mathrm{rh}<\mxt$, with $\mxo$ in the GeV range and $\mxt$ in the TeV range. For $T_\mathrm{rh}=7~\mathrm{GeV}$, the required scale lies in the range $\Lambda\simeq1-2.5~\mathrm{TeV}$, while increasing the reheating temperature to $T_\mathrm{rh}=20~\mathrm{GeV}$ shifts it to $\Lambda\simeq5-12~\mathrm{TeV}$, reflecting the UV-dominated nature of the production mechanism. The white regions correspond to points for which the required value of $\Lambda$ lies outside the displayed range.}
    \label{fig:freezeIn}
\end{figure}

The values of $\Lambda$ required to obtain $\Omega_{\xo} h^2\simeq0.12$ are shown in \cref{fig:freezeIn} for the two considered representative reheating temperatures $T_{\rm rh}=7~\mathrm{GeV}$ (left) and $T_{\rm rh}=20~\mathrm{GeV}$ (right). The qualitative behaviour follows directly from \cref{eq:freezein_lambda_estimate}. For fixed $T_{\rm rh}$ and $\mxo$ values, increasing $\mxt$ suppresses the production rate through the virtual heavy-vector exchange, so the observed abundance is recovered for a smaller effective scale. This appears as a gradual lightening of the colour map toward larger $\mxt$ values in each panel. Conversely, increasing $\mxo$ at fixed $\mxt$ and $T_{\rm rh}$ values raises the energy density associated with a given comoving abundance and therefore mildly increases the required scale, consistently with the $\Lambda_{\rm eff}\propto \mxo^{1/8}$ scaling. This appears as a slight darkening of the colour map toward larger $\mxo$ values. The dominant effect is nevertheless the sensitivity to the reheating temperature. From \cref{eq:freezein_lambda_estimate}, one has approximately $\Lambda_{\rm eff}\propto T_{\rm rh}^{11/8}$ in the hierarchy considered here. Raising the reheating temperature from $7$ to $20~\mathrm{GeV}$ therefore shifts the required scale from $\Lambda\simeq1-2.5~\mathrm{TeV}$ to $\Lambda\simeq5-12~\mathrm{TeV}$ for a given mass spectrum, as confirmed by the results in \cref{fig:freezeIn}.

These freeze-in benchmarks have direct implications for the collider phenomenology associated with the two-dark-vector model. Since the LHC production cross sections are strongly suppressed by the EFT scale, the larger values required in the $T_{\rm rh}=20~\mathrm{GeV}$ benchmark lead to rates that are much smaller than those obtained for the $\Lambda=1~\mathrm{TeV}$ collider benchmark. The $T_{\rm rh}=7~\mathrm{GeV}$ benchmark, by contrast, naturally points to scales of order a few TeV and is therefore closer to the regions of the parameter space that can be probed by current or future LHC searches for dark matter. The freeze-in contours should thus be interpreted as cosmological target regions for the EFT scale, rather than as exclusions directly comparable to collider limits.

%############################################
%############################################
%############################################
\section{LHC simulation and event selection} \label{sec:MC_CPheno}

We now describe the collider simulation used to assess the LHC sensitivity to the vector-DM signal introduced in \cref{sec:mod}. The analysis targets the prompt $\gamma+\text{jets}+\met$ topology arising from
\begin{equation}\label{eq:signalprocess}
  pp\to \xo\xt+j \quad\text{with}\quad \xt\to\xo\gamma,
\end{equation}
at a centre-of-mass energy of $\sqrt{s}=13.6~\mathrm{TeV}$. This section defines the simulation setup employed for the generation of the signal and background samples, establishes the prompt-decay and branching-fraction assumptions for the $\xt$ state and specifies the event selection used in the statistical interpretation. In \cref{subsec:samples}, we first describe the Monte Carlo event generation of the signal and SM backgrounds, thus providing details about the model implementation, the hard-scattering processes, parton showering and detector simulation. In \cref{subsec:decays_rates}, we then discuss the decay properties of the signal benchmarks, before introducing the object definitions, our baseline selection and the considered exclusive $\met$ signal regions in \cref{subsec:selection}. Finally, we examine the main kinematic features of the selected events in \cref{subsec:kinematics}.

\subsection{Event generation and simulated samples}\label{subsec:samples}
Hard-scattering signal and background samples are generated at leading order in QCD with \textsc{MadGraph5\_aMC@NLO}~\cite{Alwall:2014hca}, at a centre-of-mass energy of $\sqrt{s}=13.6~\mathrm{TeV}$. We use our own implementation of the two-vector effective theory introduced in \cref{sec:mod} in \textsc{FeynRules}~\cite{Christensen:2009jx, Alloul:2013bka}, exported in the UFO format~\cite{Degrande:2011ua, Darme:2023jdn}, and we convolute leading-order matrix elements with the \texttt{NNPDF23\_lo\_as\_0130\_qed} set of parton distribution functions~\cite{Ball:2013hta, Buckley:2014ana} for all hard-scattering simulations. At generator level we impose the following loose cuts,
\begin{equation}
  p_T^j>20~\mathrm{GeV},\qquad
  p_T^\gamma>10~\mathrm{GeV},\qquad
  |\eta_j|<5,\qquad
  |\eta_\gamma|<2.5,\qquad
  \Delta R(\gamma,j)>0.4,
\end{equation}
without any additional custom preselection. Parton-level events are next passed through \textsc{Pythia 8}~\cite{Bierlich:2022pfr} for parton showering and hadronisation, and subsequently processed with \textsc{Delphes 3} with the default ATLAS detector configuration~\cite{deFavereau:2013fsa}. In this framework, jets are reconstructed with the anti-$k_T$ algorithm~\cite{Cacciari:2008gp}, as implemented in the \textsc{FastJet} suite~\cite{Cacciari:2011ma}, while using a radius parameter $R=0.4$.

For the signal, we consider the production of an $\xt \xo$ pair of dark vectors in association with additional QCD radiation $\xo \xo j \gamma$ while also including off-shell effects, which gives rise to a prompt $\gamma+\text{jets}+\met$ final state. Our reference samples are generated for the scale $\Lambda=1~\mathrm{TeV}$ and Wilson coefficients $c_B=\tilde c_B=1$, and we scan over a grid of mass spectra satisfying $\mxt>\mxo$, consistently with $\xo$ being the stable lightest dark vector and with the radiative decay $\xt\to\xo\gamma$ being kinematically allowed. Both masses are taken below $2.5~\mathrm{TeV}$, and for each point in the mass grid $2.5\times10^4$ signal events are generated. Since the scan extends to dark-vector masses above the benchmark value $\Lambda=1~\mathrm{TeV}$, the high-mass part of the parameter space should be interpreted with care. Here, even for the lower-$\met$ signal regions of our analysis, the hard scale of the partonic process can still be comparable to or larger than the EFT suppression scale. We therefore interpret these results as benchmark EFT sensitivities rather than fully UV-complete exclusions, keeping in mind that a more conservative treatment would require either an explicit EFT truncation based on the event-by-event hard scale, or a UV completion specifying the heavy states responsible for the dimension-six operators included in the Lagrangian of \cref{eq:lagrangian}.

The leading SM backgrounds are processes with a final state featuring a prompt photon, large hadronic recoil, and either genuine missing transverse momentum or leptonic decays that can enter the signal region after reconstruction and veto requirements. We thus generate events for the processes
\begin{equation}
  pp\to Z(\to\nu\bar\nu)\gamma j,\qquad
  pp\to W(\to\ell\nu)\gamma j,\qquad
  pp\to t\bar t\gamma,
\end{equation}
where the first contribution is irreducible while the latter two enter when charged leptons $\ell=e,\mu$ are not reconstructed, fall outside the analysis acceptance or are removed by the event selection. For each background process, $3\times10^5$ events are generated. Since only a small number of $t\bar t\gamma$ events survive the final selection, its post-selection yield is affected by sizeable Monte Carlo statistical fluctuations. To avoid an overly optimistic sensitivity estimate from a possible underestimate of this subleading background, we apply a deliberately conservative rescaling factor of $10$ to the nominal $t\bar t\gamma$ yield, which should not be interpreted as a perturbative $K$-factor but rather as a robustness prescription for the prospective analysis. After this rescaling, the $t\bar t\gamma$ contribution nevertheless remains subleading with respect to the dominant $Z(\to\nu\bar\nu)\gamma j$ background. 

Other prompt-photon backgrounds, such as diboson-plus-photon production ($WZ\gamma$, $ZZ\gamma$ and related channels), as well as photon-rich final states such as $\gamma\gamma$, $W\gamma\gamma$, $Z\gamma\gamma$ and $\gamma\gamma+\text{jets}$, can in principle contribute as well. They are not included in the nominal background model of this study since they are expected to be strongly suppressed after the requirements of a large amount of missing transverse energy, a hard photon, a recoil jet, and lepton and $b$-jet vetoes. In addition, instrumental backgrounds such as $\gamma+\text{jets}$ events with fake missing energy originating from jet or photon mismeasurement are not simulated. Although these contributions are in principle important in experimental analyses, the assessment of their impact would require data-driven estimates within a full experimental analysis including photon or jet misidentification effects. Finally, given the exploratory nature of our study, all samples are normalised at leading order and no multipartonic matrix-element matching is applied. This approximation is sufficient for assessing the relative performance of the event selection and the expected sensitivity trends, but a more precise experimental reinterpretation would require matched samples and a dedicated treatment of background normalisations and higher-order corrections.

\subsection{Decay properties of the heavier dark vector} \label{subsec:decays_rates}

\begin{figure}
  \centering
  \includegraphics[width=0.48\linewidth]{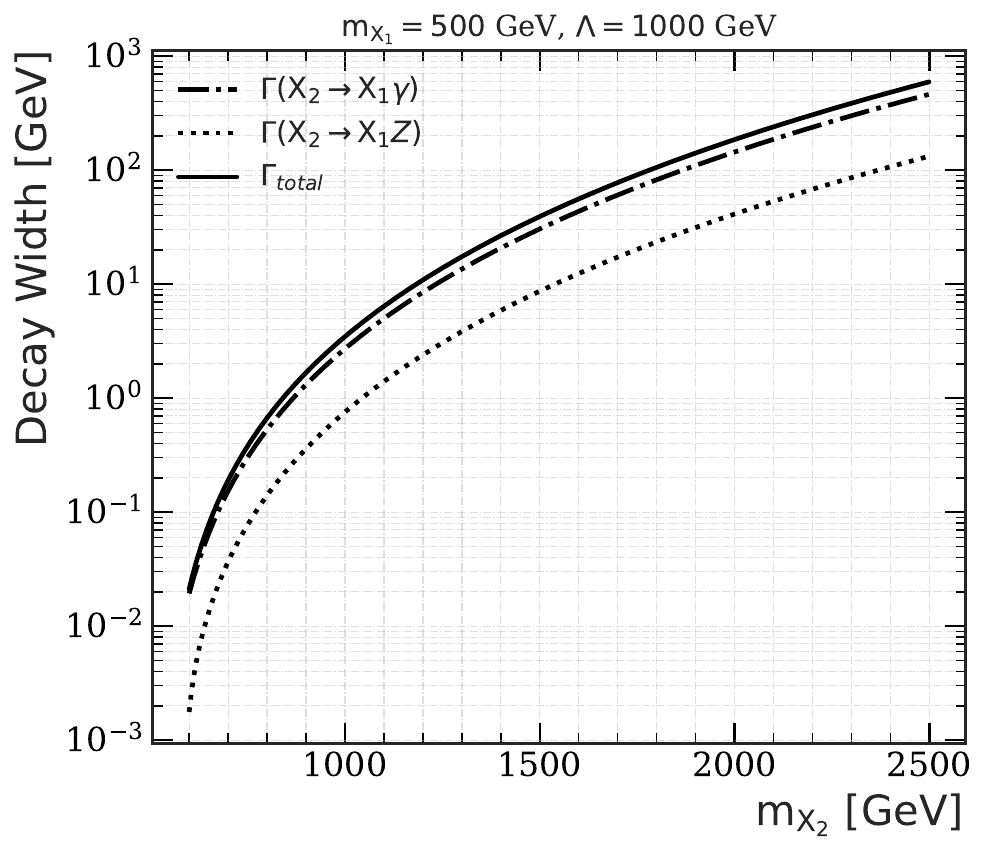}\hfill
  \includegraphics[width=0.48\linewidth]{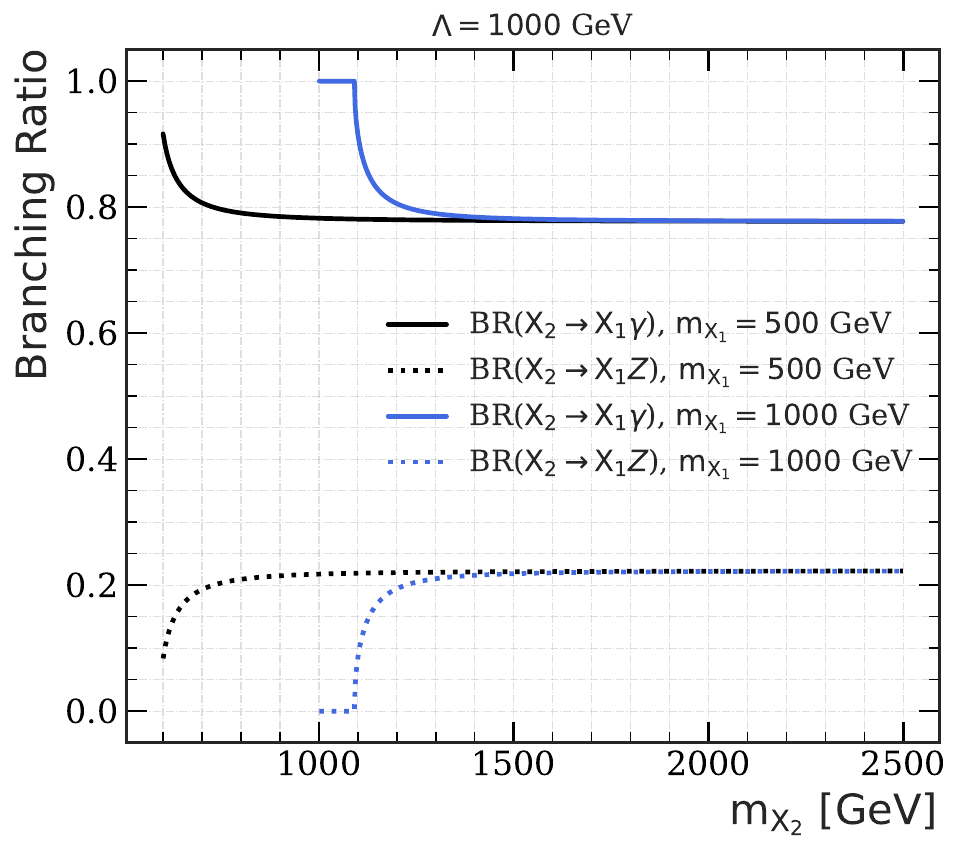}  
  \caption{Decay properties of the heavier dark vector $\xt$ for the benchmark choice $c_B=\tilde c_B=1$ and $\Lambda=1~\mathrm{TeV}$. The two-body partial widths shown in the left panel are given as functions of $\mxt$ for $\mxo=500~\mathrm{GeV}$; the dash-dotted, dotted and solid curves correspond respectively to $\Gamma(\xt\to\xo\gamma)$, $\Gamma(\xt\to\xo Z)$ and the total width. The right panel displays the corresponding branching fractions for $\mxo=500~\mathrm{GeV}$ (black) as well as for $\mxo=1~\mathrm{TeV}$ (blue), with the solid curves denoting the radiative mode and the dashed curves the $Z$-boson mode.}
  \label{fig:x2_widths}
\end{figure}

The collider signature studied in this work relies on the radiative decay of the heavier dark vector, $\xt\to\xo\gamma$. Before defining the event selection, we examine in this section the decay properties of the $\xt$ state in the benchmark setup used for the signal simulation, namely $c_B=\tilde c_B=1$ and $\Lambda=1~\mathrm{TeV}$. 

The two-body decay widths of the $\xt$ state are shown in the left panel of \cref{fig:x2_widths} as a function of $\mxt$ and for $\mxo=500~\mathrm{GeV}$. The radiative channel $\xt\to\xo\gamma$ is open throughout the displayed mass range, whereas the massive channel $\xt\to\xo Z$ becomes available only above the threshold $\mxt-\mxo>m_Z$. Both widths increase rapidly with $\mxt$, reflecting the characteristic $\mxt^5/\Lambda^4$ scaling of the dimension-six decay rates modulated by phase-space factors, as given by the closed-form expressions in \cref{eq:x2decays}. Close to threshold, the widths are suppressed by the small mass splitting. At larger mass splittings, the radiative width remains the dominant contribution, owing to the larger hypercharge projection onto the photon and to the residual phase-space suppression of the massive $Z$ channel, while the $Z$-boson mode becomes a non-negligible component of the total width once it is kinematically open. For $\mxt\gtrsim1~\mathrm{TeV}$, however, the total width reaches $\Gamma_\mathrm{tot}/\mxt\sim\mathcal{O}(\mathrm{few}\%)$ and grows to $\sim20\%$ at $\mxt=2.5~\mathrm{TeV}$, at which point the narrow-width approximation underlying the on-shell signal simulation becomes questionable. This coincides with the regime where the hard scales involved in producing the dark-vector pair can exceed the benchmark EFT suppression scale, and results in this region should therefore be interpreted with caution. 

The corresponding branching fractions are displayed in the right panel of \cref{fig:x2_widths} for two representative DM mass choices, $\mxo=500~\mathrm{GeV}$ and $\mxo=1~\mathrm{TeV}$. For each value of $\mxo$, the radiative branching ratio is equal to unity below the on-shell $Z$ threshold (which lies below the plotted range for the $\mxo=500~\mathrm{GeV}$ benchmark), since $\xt\to\xo\gamma$ is then the only available two-body electroweak decay mode. Once the $Z$ channel opens, the quantity ${\rm BR}(\xt\to\xo\gamma)$ decreases. Nevertheless, the radiative branching fraction remains large over the mass range relevant for the collider analysis. In addition, at large mass splittings where the $Z$-boson mass and threshold effects become negligible, the $c_B^2$ and $\tilde c_B^2$ contributions to the $Z$-channel width approach the same kinematic dependence as in the radiative channel, up to the replacement $c_W^2\to s_W^2$. The two branching fractions therefore tend towards the pattern inherited from the electroweak decomposition of the hypercharge field strength, $B_{\mu\nu}=c_W A_{\mu\nu}-s_W Z_{\mu\nu}$. In this limit the photon and $Z$ partial widths are then approximately weighted by $c_W^2$ and $s_W^2$, respectively, so that the radiative branching fraction tends toward $c_W^2\simeq0.77$, while the $Z$-boson branching fraction tends toward $s_W^2\simeq0.23$, although the convergence is gradual and the asymptotic values are not fully reached within the displayed mass range.

\begin{figure}
  \centering
  \includegraphics[width=0.48\linewidth]{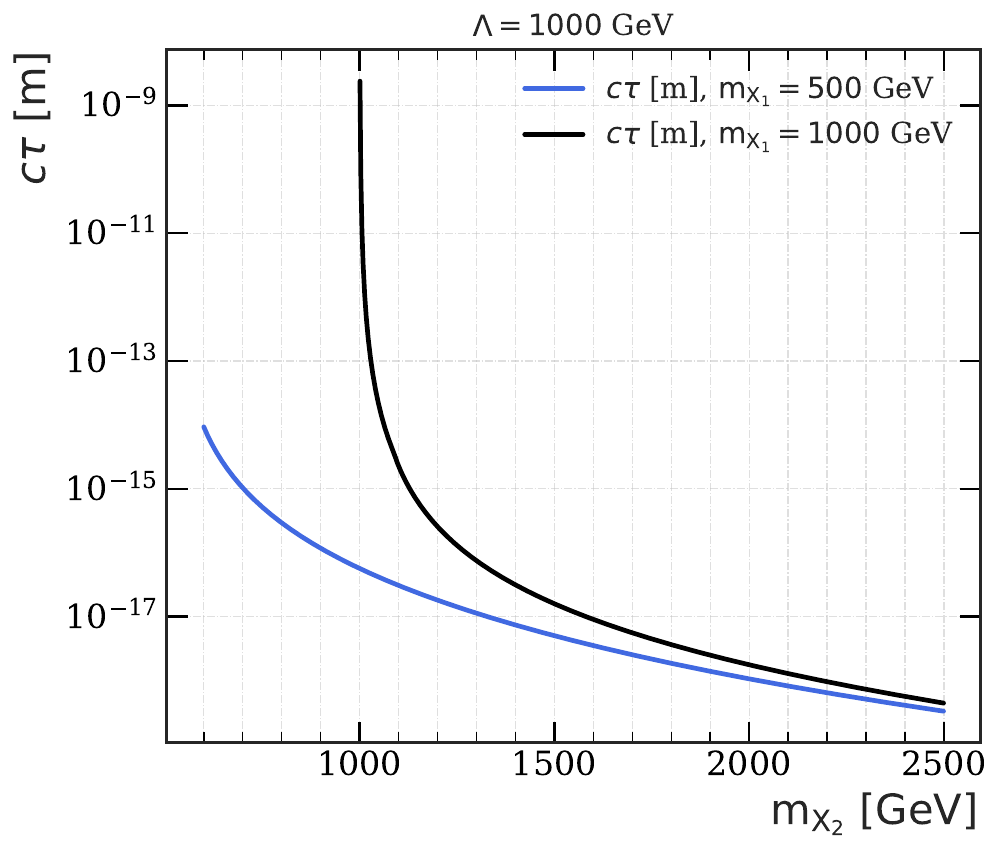}
  \caption{Proper decay length of the heavier dark vector $\xt$ as a function of $\mxt$ for the benchmark choice $c_B=\tilde c_B=1$ and $\Lambda=1~\mathrm{TeV}$. The two curves correspond to $\mxo=500~\mathrm{GeV}$ (blue) and $\mxo=1~\mathrm{TeV}$ (black).}
  \label{fig:x2_ctau}
\end{figure}

At the same time, the $\xt$ decay remains prompt throughout the displayed parameter range. This is confirmed quantitatively in \cref{fig:x2_ctau}, which shows the proper decay length $c\tau$ as a function of $\mxt$ for both considered mass benchmarks. The decay length is the largest near the kinematic threshold, where the phase-space factor $(1-\varrho)^3$ in the radiative decay width strongly suppresses the total width. For $\mxo=1~\mathrm{TeV}$ close to threshold, one finds $c\tau\sim\mathcal{O}(1~\mathrm{nm})$, while for $\mxo=500~\mathrm{GeV}$ the decay length at the left edge of the displayed range is already microscopic, of $\mathcal{O}(10~\mathrm{fm})$. In both cases, $c\tau$ decreases rapidly as $\mxt$ increases, tracking the $\mxt^5/\Lambda^4$ growth of the total width. Even at the maximum displayed value, the lab-frame displacement $L\sim\gamma\beta\,c\tau$ remains at most at the nanometre scale for the boosts relevant to the selected signal events, well below the micrometre-scale spatial resolution of LHC inner tracking detectors. The prompt-decay assumption used in the signal simulation is therefore well justified across the entire parameter space considered here. Departures from this regime would require either a substantially larger EFT scale or an extremely compressed spectrum. Starting from the largest decay lengths shown in \cref{fig:x2_ctau}, reaching the displaced-vertex regime, $c\tau\sim\mathrm{mm}$, would indeed require increasing the EFT scale by roughly a factor of $30$ since $c\tau\propto\Lambda^4$, or a mass splitting well below the GeV scale where the $(1-\varrho)^3$ suppression renders the radiative width very small. Such configurations are outside the scope of the present collider analysis but could motivate dedicated long-lived-particle studies.

%############################################
%############################################
%############################################

\subsection{Object definitions, baseline selection and signal regions} \label{subsec:selection}
The event selection is designed to isolate the prompt $\gamma+\text{jets}+\met$ topology of the signal while keeping the analysis sufficiently simple for a prospective sensitivity study. It is inspired by the ATLAS search for final states with photons, jets and missing transverse momentum of Ref.~\cite{ATLAS:2022ckd}, adapted to the signal topology considered here, rather than by a dedicated multivariate or multidimensional optimisation. The targeted final state contains a hard isolated photon from the radiative decay $\xt\to\xo\gamma$, missing transverse momentum from the two invisible $\xo$ particles, and at least one recoil jet from QCD radiation. The main backgrounds are reduced by exploiting the properties of these objects, together with angular requirements that suppress configurations in which the reconstructed missing transverse momentum is aligned with a visible object. Lepton and $b$-jet vetoes are additionally used to reduce backgrounds with leptonic vector-boson decays and top-quark production, in particular originating from the $W\gamma j$ and $t\bar t\gamma$ processes.

Our baseline selection requires events to contain at least one photon satisfying the photon-identification and isolation criteria implemented in the default ATLAS \textsc{Delphes} card, with
\begin{equation}
  p_T^\gamma>150~\mathrm{GeV} \qquad\text{and}\qquad |\eta^\gamma|<2.37.
\end{equation}
This photon requirement is directly motivated by the two-body decay kinematics of the $\xt$ dark vector. In the $\xt$ rest frame, the photon indeed carries an energy $E_\gamma^\star=(\mxt^2-\mxo^2)/(2\mxt)$, so TeV-scale spectra with sizeable mass splittings naturally lead to energetic photons with transverse momenta that can reach several hundred GeV after boosting to the laboratory frame. Events must also contain at least one reconstructed jet satisfying
\begin{equation}
  p_T^j>100~\mathrm{GeV} \qquad\text{and}\qquad |\eta^j|<2.5,
\end{equation}
with the photon-jet ambiguities being treated through the standard \textsc{Delphes} object-uniqueness procedure in which the reconstructed photons take priority over the reconstructed jets. We further select events containing significant missing transverse momentum,
\begin{equation}\label{eq:metcut}
  \met \ge 200~\mathrm{GeV},
\end{equation}
which exploits the recoil of the two invisible $\xo$ particles against the photon and QCD radiation in the signal topology. The leading photon and leading jet are then used to impose
\begin{equation}
  \Delta\phi(j_1,\met)>0.4
  \qquad\text{and}\qquad
  \Delta\phi(\gamma_1,\met)>0.4.
\end{equation}
These angular requirements reject events in which the missing transverse momentum is aligned with the leading visible objects, as expected for backgrounds affected by jet or photon mismeasurement. Finally, events containing reconstructed isolated electrons or muons with $p_T>10~\mathrm{GeV}$, or any $b$-tagged jet, are vetoed.

\begin{table}
  \centering\renewcommand{\arraystretch}{1.4} \setlength{\tabcolsep}{9pt}
  \begin{tabular}{l | cccc}
    \toprule
    Selection requirement & single-SR & $\mathrm{SR}_{1}$ & $\mathrm{SR}_{2}$ & $\mathrm{SR}_{3}$\\
    \midrule
    Photon  & \multicolumn{4}{c}{$p_T^\gamma>150~\mathrm{GeV}$, $|\eta^\gamma|<2.37$} \\
    Jet  & \multicolumn{4}{c}{$p_T^j>100~\mathrm{GeV}$, $|\eta^j|<2.5$} \\
    Angular separation & \multicolumn{4}{c}{$\Delta\phi(j_1,\met)>0.4$, $\Delta\phi(\gamma_1,\met)>0.4$} \\
    Lepton veto & \multicolumn{4}{c}{No isolated electrons or muons with $p_T>10~\mathrm{GeV}$} \\
    $b$-jet veto & \multicolumn{4}{c}{Reject events containing any $b$-tagged jet} \\
    \midrule
    $\met$ requirement & $\ge200~\mathrm{GeV}$ & $\in [200, 350[~\mathrm{GeV}$ & $\in [350, 450[~\mathrm{GeV}$ & $\ge450~\mathrm{GeV}$ \\
    \bottomrule
  \end{tabular}
  \caption{\label{tab:Cuts} Summary of the baseline selection and signal-region definitions. The single-SR configuration corresponds to an inclusive counting region with $\met\ge200~\mathrm{GeV}$, while the three-SR configuration partitions the same baseline-selected events into three mutually exclusive $\met$ intervals.}
\end{table}

\begin{table}
  \centering\renewcommand{\arraystretch}{1.4} \setlength{\tabcolsep}{12pt}
  \begin{tabular}{l | ccccc}
  \toprule
    Process & $\sigma_{\rm LO}$ [pb] & single-SR & $\mathrm{SR}_1$ & $\mathrm{SR}_2$ & $\mathrm{SR}_3$\\
    \midrule
    $W\gamma j$, $W\to \ell\nu$ & 33.85 & 188  & 141 & 31 &  16 \\
    $Z\gamma j$, $Z\to \nu\bar{\nu}$ & 0.69 & 103 & 73 & 19 & 12 \\
    $t\bar{t}\gamma$ & 6.9  & 406 & 371 &  32  &  4 \\
   \bottomrule
  \end{tabular}
  \caption{Expected SM background yields after the common baseline selection and after the exclusive $\met$ binning, for $\mathcal{L}=139~\mathrm{fb}^{-1}$. The cross sections $\sigma_{\rm LO}$ correspond to the leading-order generation rates used for normalisation, and the $t\bar t\gamma$ yields include the conservative rescaling factor of $10$ discussed in the text.}
  \label{tab:BGs}
\end{table}

After applying this baseline analysis strategy, we consider two alternative ways of using the selected events for statistical interpretation by defining signal regions corresponding to either an inclusive or a binned treatment of the $\met$ requirement. We first define a single inclusive signal region, denoted `single-SR', in which all events satisfying the baseline selection are collected. These events therefore obey the generic missing-transverse-momentum requirement of \cref{eq:metcut}, $\met \ge 200~\mathrm{GeV}$. We next consider a three-bin configuration, denoted `three-SR', in which the same baseline-selected events are partitioned into the mutually exclusive $\met$ intervals
\begin{equation}\begin{split}
  \mathrm{SR}_{1}:&\quad 200 \le \met < 350~\mathrm{GeV}, \\[.1cm]
  \mathrm{SR}_{2}:&\quad 350 \le \met < 450~\mathrm{GeV}, \\[.1cm]
  \mathrm{SR}_{3}:&\quad \met \ge 450~\mathrm{GeV}. 
\end{split}\end{equation}
The bin boundaries are chosen to retain coarse information on the $\met$ spectrum while keeping sufficient simulated background statistics in each region. We emphasise that this three-SR configuration does not introduce any additional object-level requirements, and therefore only retains the associated coarse shape information which is later exploited in a binned likelihood analysis. The complete selection is summarised in \cref{tab:Cuts} and the resulting expected background yields for $\mathcal{L}=139~\mathrm{fb}^{-1}$ are reported in \cref{tab:BGs}. The table illustrates the migration of the background composition across the $\met$ regions: after the conservative rescaling of the $t\bar t\gamma$ contribution, $t\bar t\gamma$ production is sizeable in the inclusive and low-$\met$ regions, while the higher-$\met$ bins are increasingly dominated by the irreducible $Z(\to\nu\bar\nu)\gamma j$ background.

\begin{figure}
  \centering
  \includegraphics[width=0.48\linewidth]{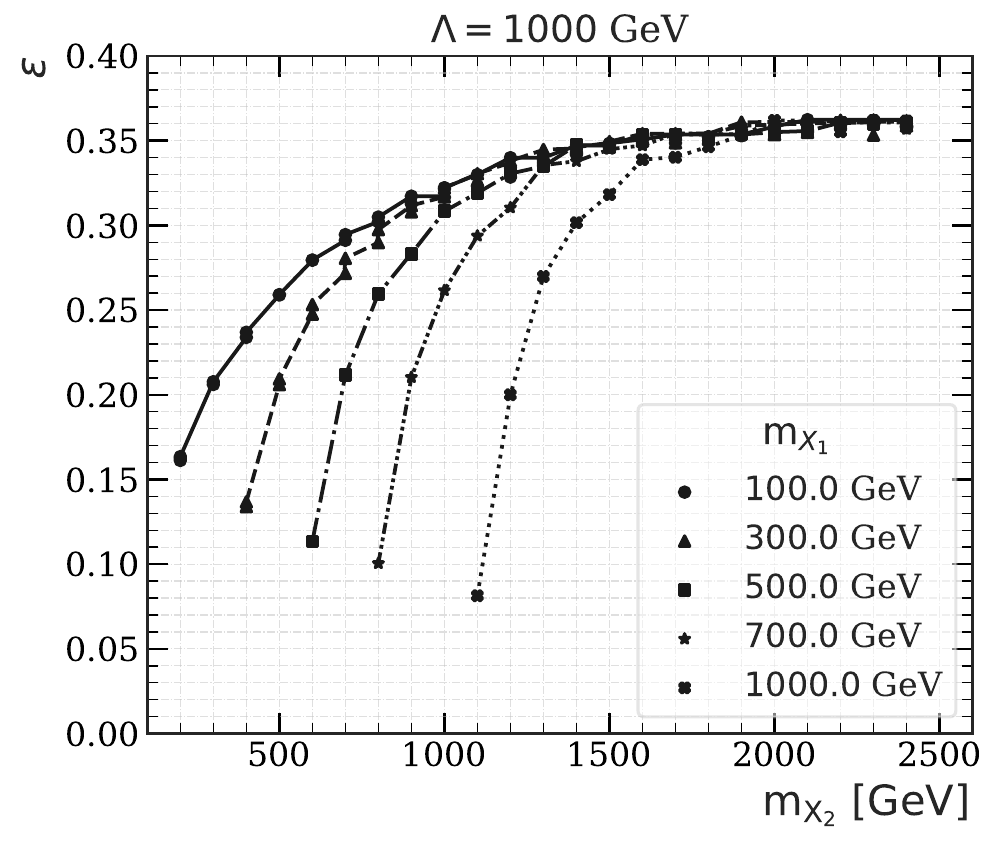}\hfill
  \includegraphics[width=0.48\linewidth]{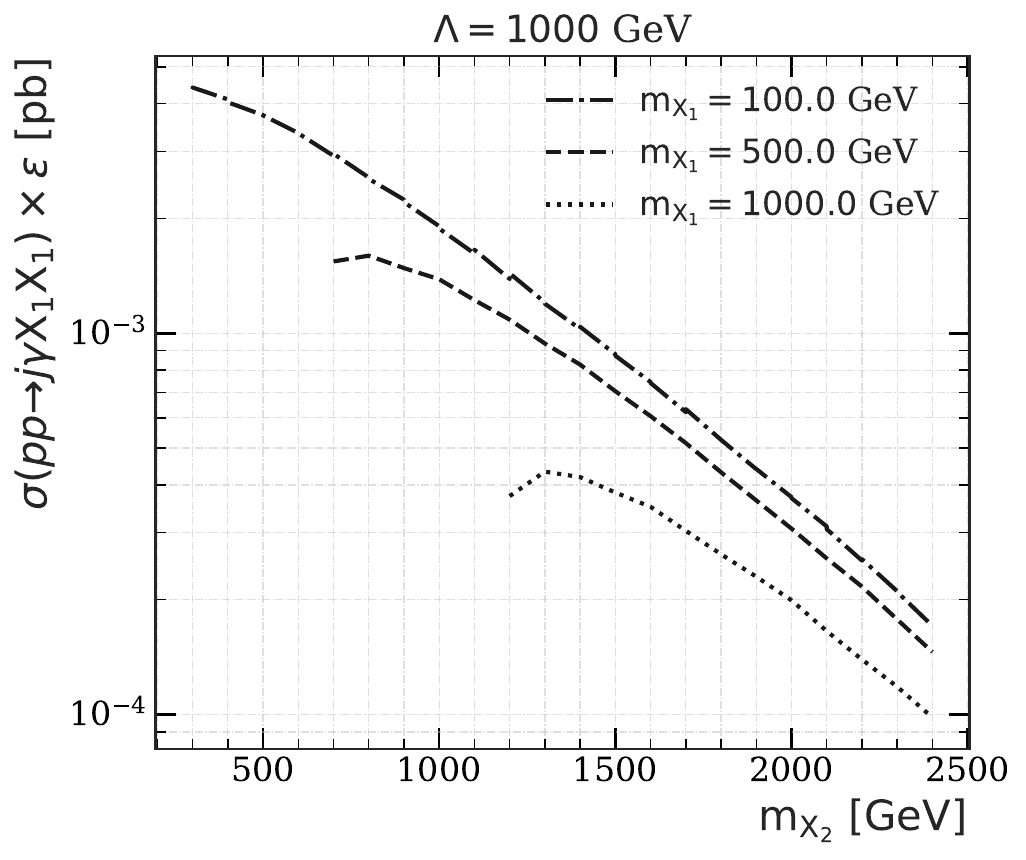}
  \caption{Baseline-selection efficiency (left) and fiducial signal rate (right) for representative signal benchmarks, with $c_B=\tilde c_B=1$ and $\Lambda=1~\mathrm{TeV}$.}
  \label{fig:baseline_eff_rate}
\end{figure}

The impact of the baseline selection on representative signal benchmark configurations is shown in \cref{fig:baseline_eff_rate}. The left panel of the figure displays the selection efficiency $\varepsilon$ as a function of the heavy dark vector mass $\mxt$ for several values of the DM mass $\mxo$. For a fixed $\mxo$ value, the efficiency increases with $\mxt$, reflecting the larger mass splitting and hence the harder photon produced in the two-body decay $\xt\to\xo\gamma$. The missing-transverse-momentum spectrum is also hardened as the invisible system recoils against the photon and QCD radiation. Close to the compressed regime, by contrast, the photon energy in the $\xt$ rest frame $E_\gamma^\star$ is reduced, and a larger fraction of events fails the photon and $\met$ requirements. At larger dark vector mass $\mxt$, the efficiency approaches a plateau of about $0.35-0.36$ regardless of the precise value of the DM mass, indicating that the baseline kinematic thresholds are no longer the dominant limitation. The residual losses are then expected to be driven primarily by the finite photon acceptance and the identification/isolation efficiency, with additional contributions from the jet acceptance, the angular requirements and the event vetoes.

The right panel of \cref{fig:baseline_eff_rate} shows the corresponding fiducial signal rate after the baseline selection. This quantity therefore reflects three effects at once: the production rate, the probability for the heavier dark vector to decay into the photon channel and the baseline-selection efficiency. For fixed masses and Wilson coefficients, the production cross section scales as $\sigma(pp\to\xo\xt+j)\propto\Lambda^{-4}$, since the hard production amplitude contains a single insertion of the dimension-six operator included in the Lagrangian of \cref{eq:lagrangian}, and the branching fractions and the baseline efficiency are essentially independent of $\Lambda$ as long as the decay remains prompt and the event kinematics are unchanged. The fiducial rates shown here could thus be approximately rescaled by $\Lambda^{-4}$ for other choices of the EFT scale. This rescaling should however be interpreted with the EFT-validity caveat discussed above. For mass points or event kinematics probing hard scales comparable to or larger than $\Lambda$, the contact-operator description is indeed only a benchmark parametrisation rather than a fully reliable UV-complete prediction. 

The opening of the competing $Z$ channel $\xt\to\xo Z$ reduces the branching ratio $\mathrm{BR}(\xt\to\xo\gamma)$ and therefore suppresses the visible photon rate. This effect is most relevant for the $\mxo=1~\mathrm{TeV}$ and $500~\mathrm{GeV}$ benchmarks, for which the $Z$ threshold lies within the displayed $\mxt$-range. For the $\mxo=100$ benchmark the $Z$ threshold instead lies below the left edge of the plotted range, so the branching-ratio variation is not visible in the displayed curve.

In general, the fiducial rate is found to decrease as the dark vector mass $\mxt$ increases. This fall is driven by the production cross section, since heavier dark-vector pairs require larger partonic centre-of-mass energies for which the parton luminosities are strongly suppressed. The only visible exceptions to this monotonic decrease occurs for the $\mxo=1~\mathrm{TeV}$ and $500~\mathrm{GeV}$ benchmarks close to threshold. In these regions, the production cross section is still relatively large, but the baseline efficiency is strongly suppressed because the spectrum is compressed and the photon and $\met$ requirements are difficult to satisfy. As $\mxt$ increases, the efficiency rises faster than the production cross section falls, leading to a local maximum in $\sigma\times\varepsilon$. On the other hand, for a fixed $\mxt$ value, increasing the DM mass $\mxo$ also reduces the rate, both by increasing the mass of the invisible final-state system and, for smaller mass splittings, by softening the photon. The expected sensitivity is therefore governed by the competition between an efficiency that improves away from compressed spectra and a production rate that falls steeply at high masses.

%############################################
%############################################
%############################################
\subsection{Signal and background kinematics} \label{subsec:kinematics}

\begin{figure}
  \centering
  \includegraphics[width=0.485\linewidth]{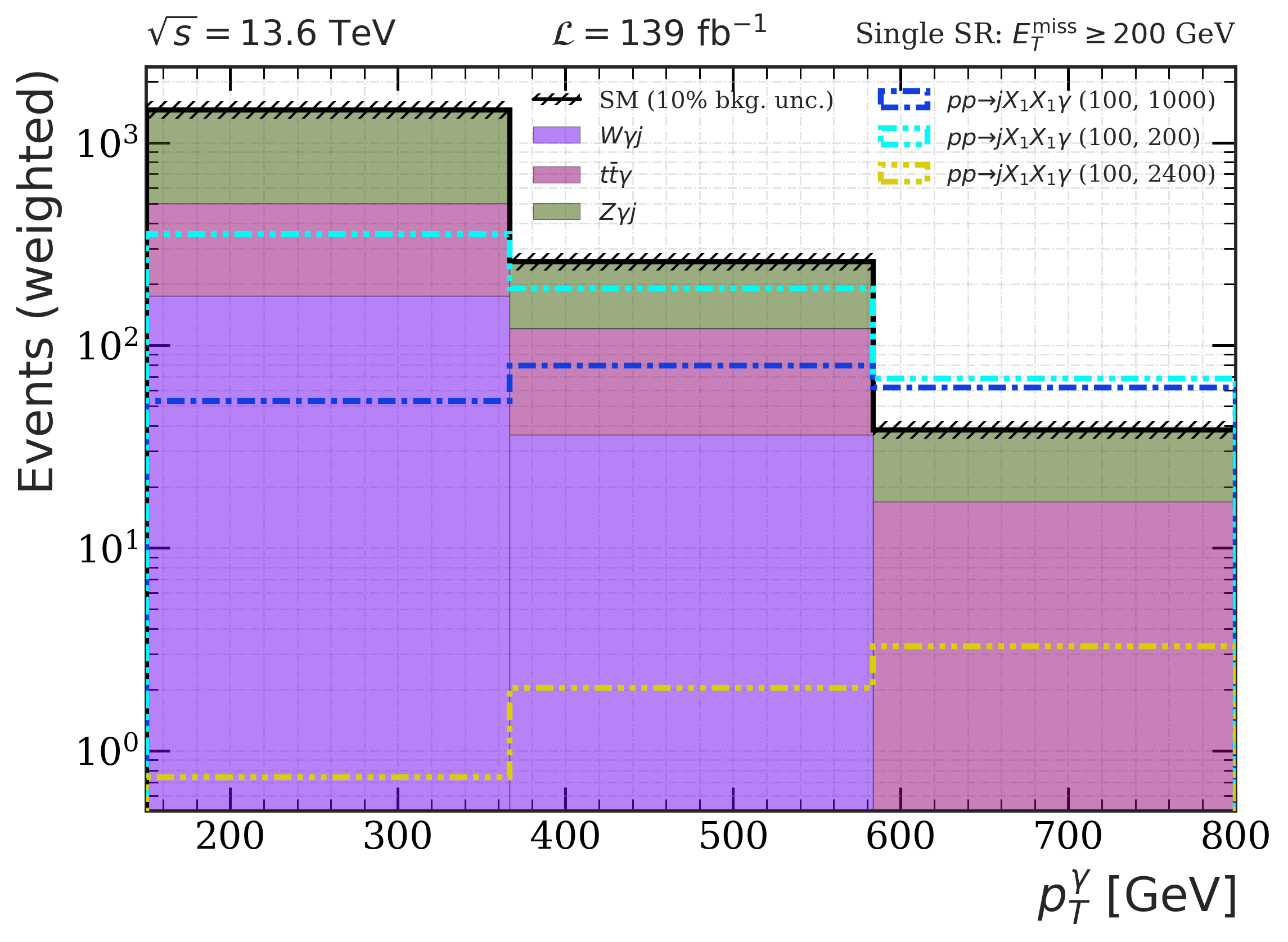}\hfill
  \includegraphics[width=0.485\linewidth]{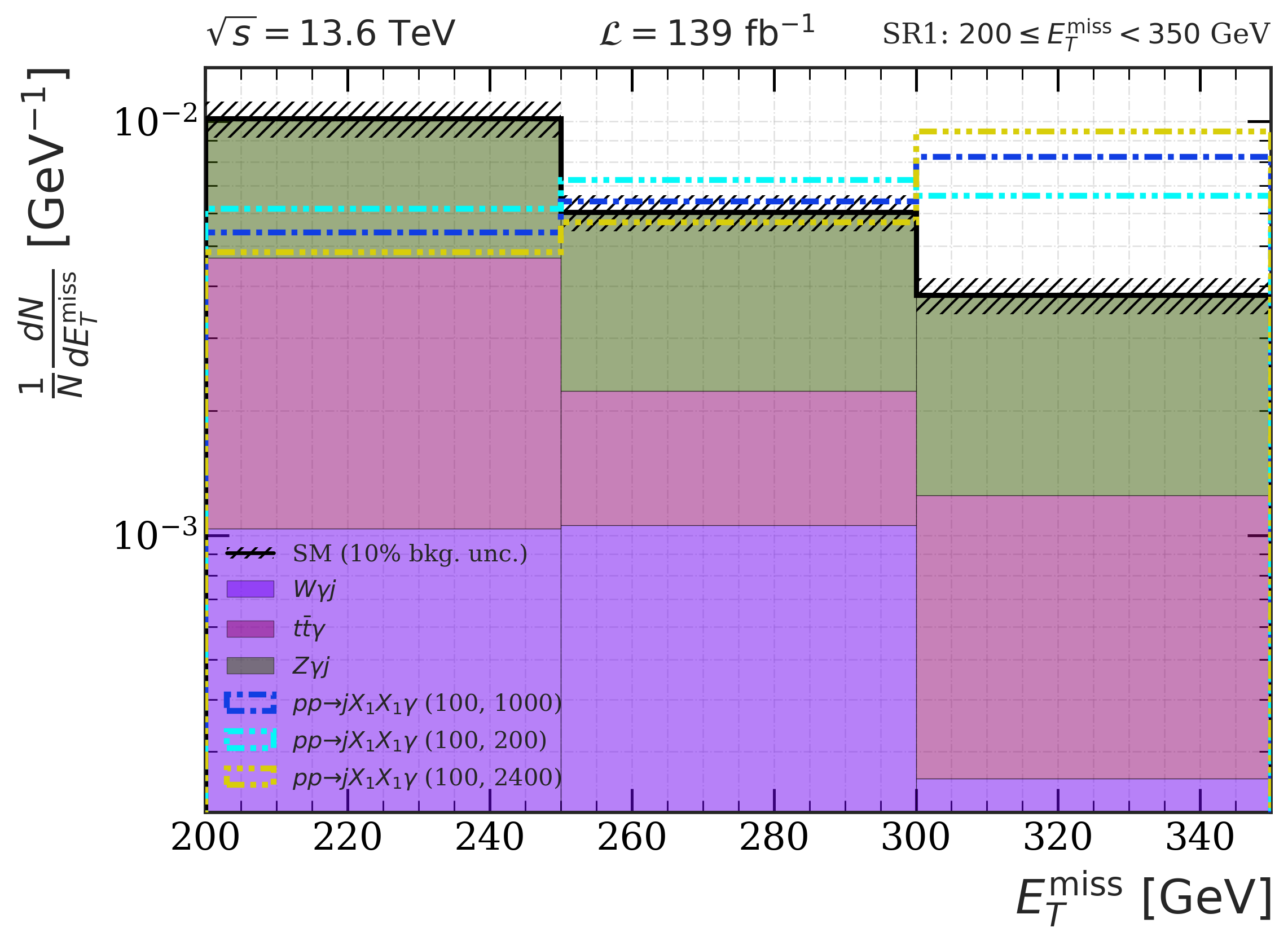}\\[.4cm]
  \includegraphics[width=0.485\linewidth]{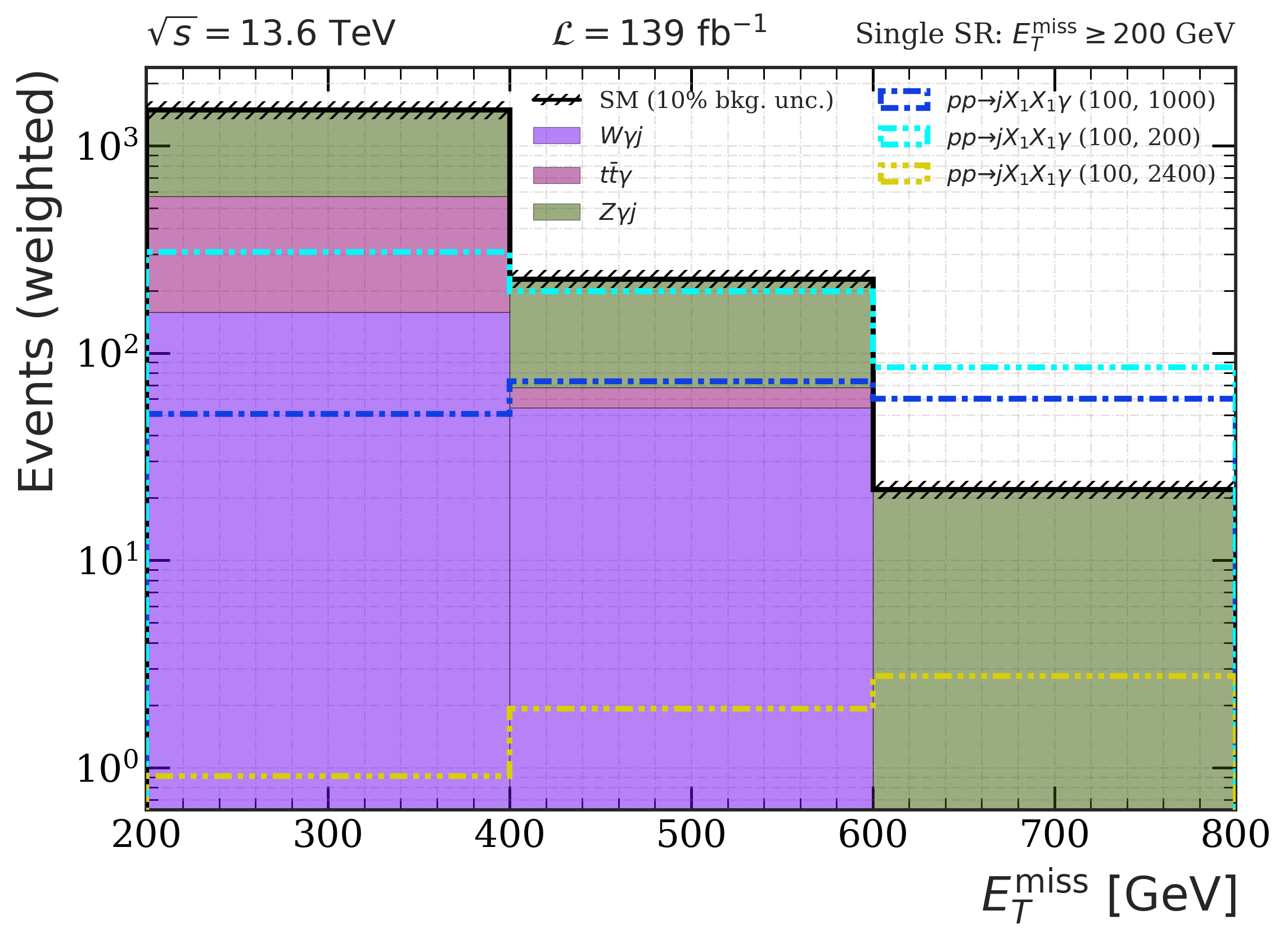}\hfill
  \includegraphics[width=0.485\linewidth]{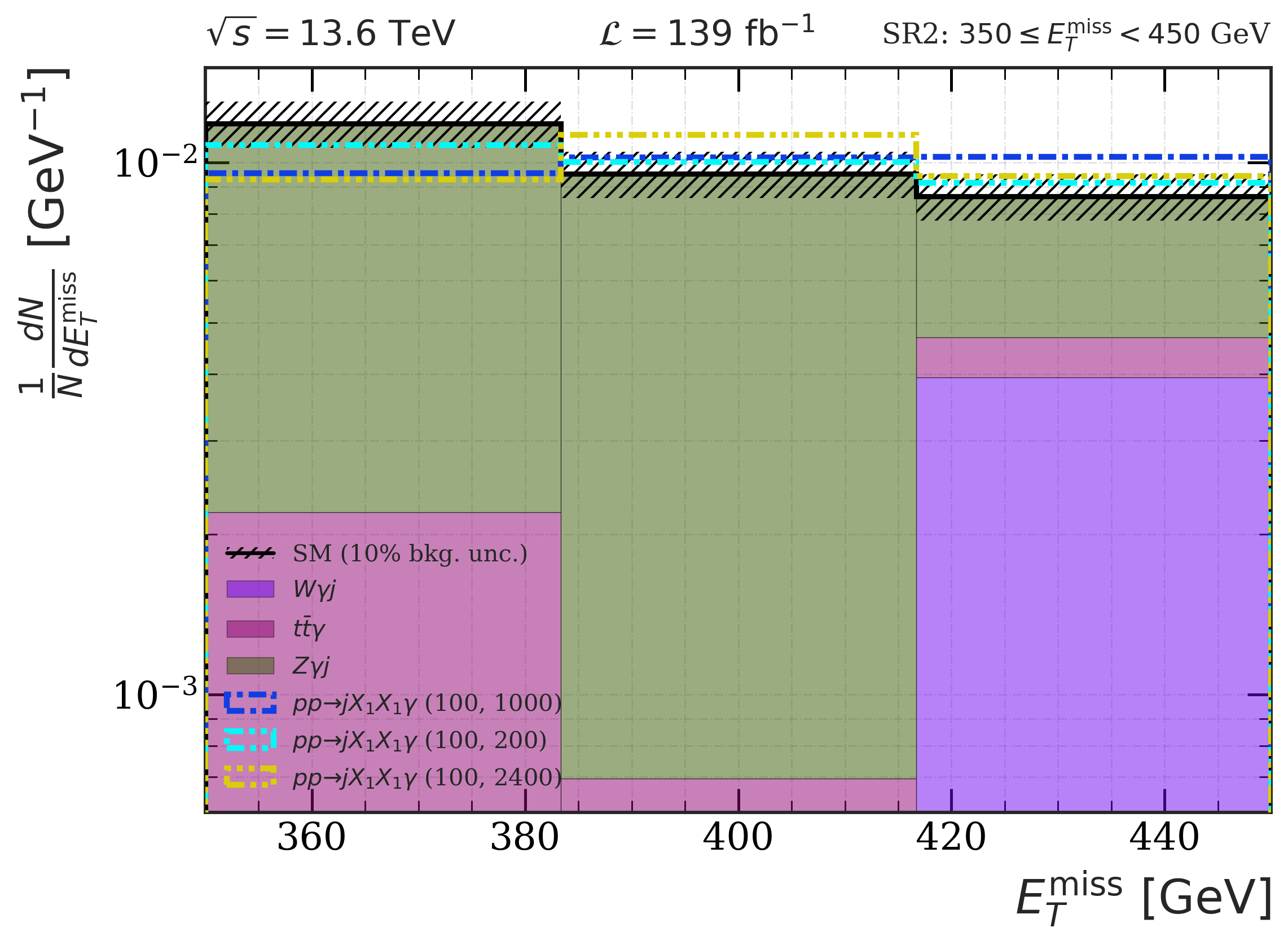}\\[.4cm]
  \includegraphics[width=0.485\linewidth]{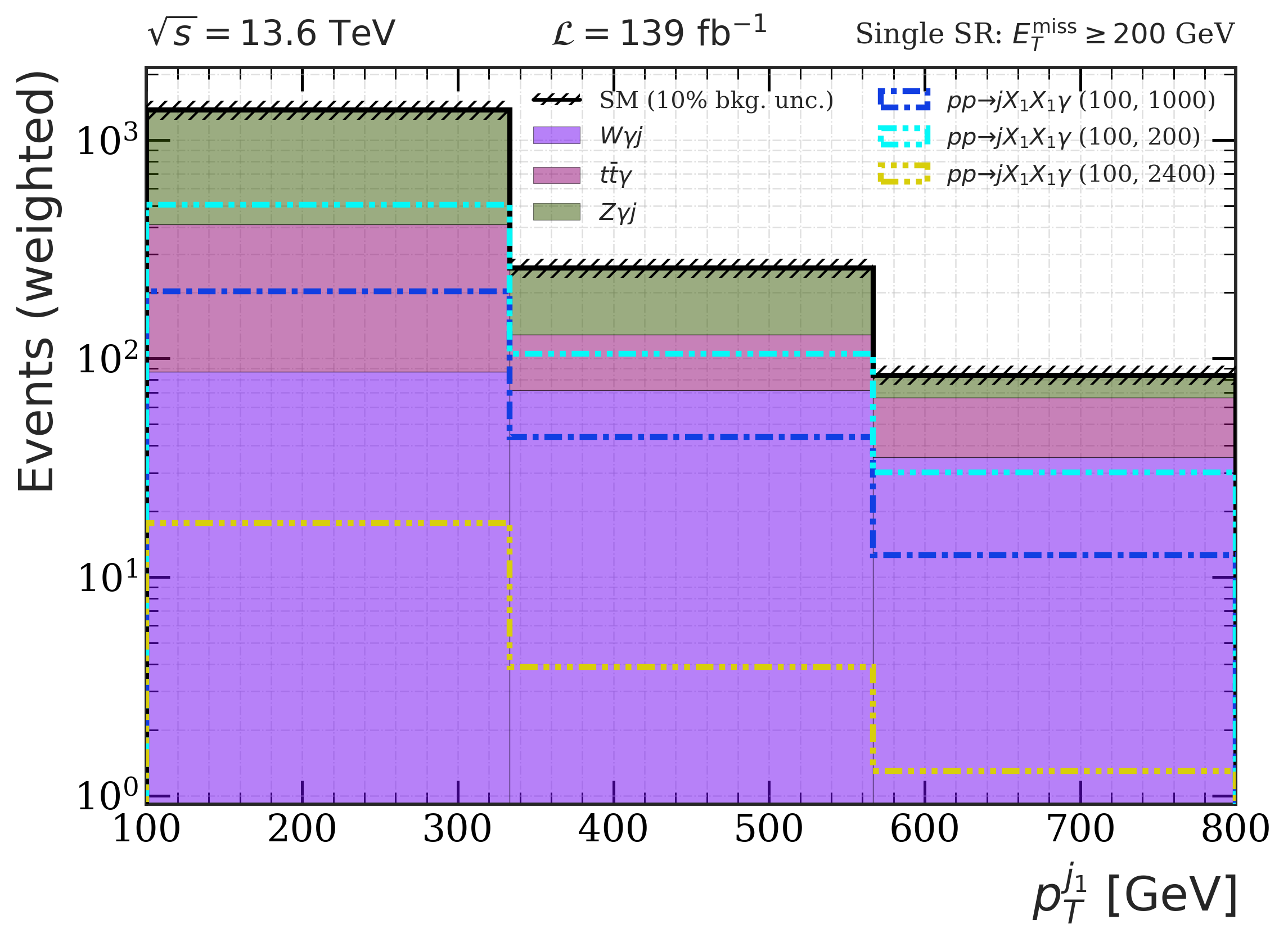}\hfill
  \includegraphics[width=0.485\linewidth]{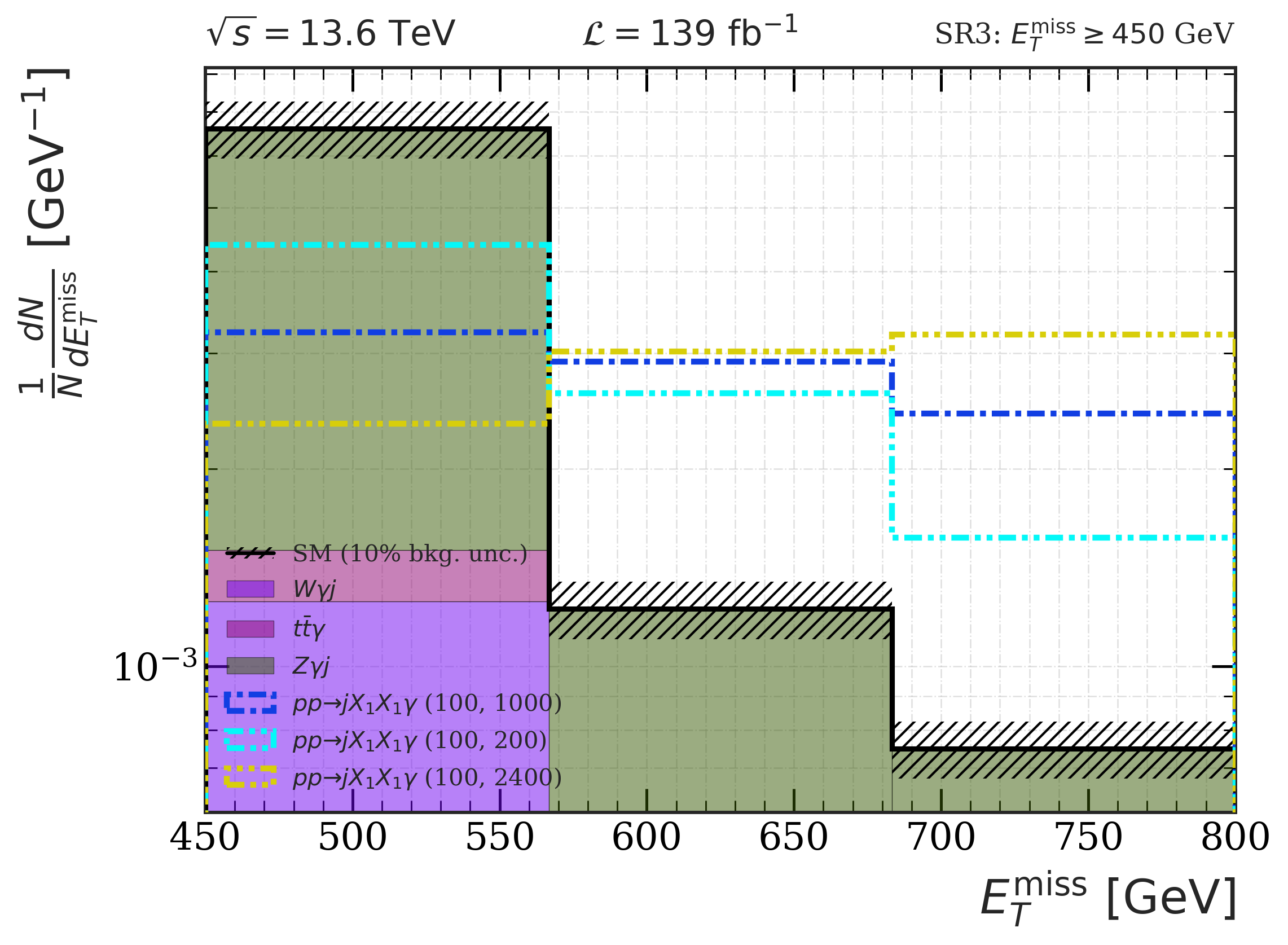}\\[.2cm]
  \caption{
  Kinematic properties of the selected events for $\mathcal{L}=139~\mathrm{fb}^{-1}$ at $\sqrt{s}=13.6~\mathrm{TeV}$. The left column shows the weighted single-SR distributions of the photon transverse momentum (top), missing transverse momentum (centre) and leading-jet transverse momentum (bottom) after the baseline selection. The stacked histograms correspond to the simulated SM backgrounds, with the hatched band including a $10\%$ systematic uncertainty on the total background yield. Representative signal benchmarks are overlaid and labelled by $(\mxo,\mxt)$ in GeV. The right column shows the normalised $\met$ distributions in the exclusive $\mathrm{SR}_1$ (top), $\mathrm{SR}_2$ (centre) and $\mathrm{SR}_3$ (bottom) signal regions, highlighting the shape information retained by the binned analysis.}
  \label{fig:kinematic_distributions}
\end{figure}

We now examine the kinematic properties of the events passing the baseline selection in order to provide useful insight into the variables driving the sensitivity of our analysis and motivate the use of the $\met$-binned signal regions introduced above. In the left column of \cref{fig:kinematic_distributions}, we show the photon transverse momentum, the missing transverse momentum and the leading-jet transverse momentum in the inclusive single-SR configuration, for an integrated luminosity of $139~\mathrm{fb}^{-1}$. The stacked histograms correspond to the simulated SM backgrounds, while the hatched band includes a $10\%$ systematic uncertainty assigned to the total background yield. Three representative signal benchmarks are overlaid, labelled by the pair of mass values $(\mxo, \mxt)$ in GeV. 

The photon transverse-momentum distribution (top left panel of the figure) directly reflects the two-body decay $\xt\to\xo\gamma$ and the rapidly falling signal production rate at high masses. In the $\xt$ rest frame, the photon energy is fixed by $E_\gamma^\star=(\mxt^2-\mxo^2)/(2\mxt)$, so larger mass splittings tend to produce harder photons and the associated benchmarks populate the high-$p_T^\gamma$ tail more efficiently. Conversely, for compressed spectra such as the benchmark defined by $(\mxo,\mxt)=(100,200)~\mathrm{GeV}$, one has lower $E_\gamma^\star$ values so that events passing the hard-photon requirement rely significantly on the boost of the decaying $\xt$ state and on the recoil against QCD radiation. At the same time, the visible signal yield is controlled by the corresponding production cross section and the radiative branching fraction. Consequently, the light benchmark $(\mxo,\mxt)=(100,200)~\mathrm{GeV}$ can still give a large event yield despite its smaller mass splitting, whereas the much heavier $(100,2400)~\mathrm{GeV}$ benchmark has a much smaller rate even though its decay photon is kinematically hard. 

The missing-transverse-momentum distribution (central left panel of the figure) is particularly important for the analysis, as it is shaped by the recoil of the dark-vector system against the photon and QCD radiation. In the figure, the displayed range is chosen to cover the kinematic region relevant for the signal-region definitions and for which the simulated background samples provide statistically stable predictions, with the associated uncertainties being included in the statistical interpretation in \cref{sec:stats}. The not-shown signal high-energy tails of the $\met$ spectrum should nevertheless be interpreted with the EFT-validity caveat discussed earlier, since harder partonic configurations are preferentially selected by the analysis. In this context, we have checked from our simulations that the signal distributions fall smoothly beyond the displayed range, both for the missing transverse momentum and for the other kinematic observables considered.

In the missing-transverse-momentum spectrum, the SM backgrounds fall steeply with increasing $\met$ values, as expected for $V\gamma j$ (with $V=W, Z$) and $t\bar t\gamma$ production after the baseline selection. The composition of the SM background however changes across the spectrum. In the lowest-$\met$ region, reducible backgrounds originating from $W\gamma j$ and $t\bar t\gamma$ production still provide sizeable contributions when the final-state charged leptons are not reconstructed or fall outside the acceptance. These components decrease rapidly with increasing $\met$ values, and the $t\bar t\gamma$ contribution becomes even nearly absent in the intermediate-$\met$ and high-$\met$ regions. As a result, these regions are increasingly dominated by the irreducible $Z(\to\nu\bar\nu)\gamma j$ background. In particular, the signal region $\mathrm{SR}_3$ is effectively controlled by the hard $\met$ tail of $Z\gamma j$ production, so the sensitivity in this region is governed by the competition between the irreducible $Z\gamma j$ background and the hard recoil spectrum of the signal. Here, several signal benchmarks exhibit a harder $\met$ spectrum than the SM backgrounds, thus indicating that the high-$\met$ part of the region $\mathrm{SR}_3$ is where the signal can differ most strongly from the background. This behaviour is exactly the one that motivates the use of exclusive $\met$ signal regions rather than a single inclusive event count. 

By contrast, the leading-jet transverse momentum $p_T^{j_1}$ (bottom left panel of the figure) is less directly tied to the dark-sector mass splitting. The jet mainly originates from QCD radiation recoiling against the electroweakly produced dark-vector pair, and similar recoil configurations are also present in the dominant SM backgrounds. The leading-jet spectrum is therefore useful for background rejection through the baseline requirements, but it is less discriminating than the missing transverse momentum. This supports the choice of using the missing transverse energy rather than $p_T^{j_1}$ as the variable defining the exclusive signal regions. 

To isolate the shape information from the overall normalisation, the right column of \cref{fig:kinematic_distributions} shows the normalised $\met$ distributions in the three exclusive signal regions, which help visualise how the selected signal and background events populate each $\met$ interval once the overall rate has been factored out. In the $\mathrm{SR}_1$ region corresponding to $200\le\met<350~\mathrm{GeV}$, sizeable shape differences are already visible for the considered benchmarks, reflecting the fact that the signal and background populate the low-$\met$ interval differently even close to the baseline threshold. In the signal region $\mathrm{SR}_2$, the normalised shapes are more similar, so this intermediate region mainly provides an additional yield bin rather than a particularly distinctive shape handle. In the last region $\mathrm{SR}_3$ where $\met\ge450~\mathrm{GeV}$, the comparison again becomes more discriminating: the SM background is concentrated toward the lower part of the interval, whereas some signal benchmarks retain a broader support at larger $\met$. 

The obtained global pattern thus explains why the three-SR strategy can improve upon a single inclusive counting region. The low-$\met$ region retains statistical power while the higher-$\met$ regions exploit the harder signal spectrum and the reduced background rate. However, for simplicity, the statistical interpretation based on the binned likelihood introduced in \cref{sec:stats} does not rely on the detailed shape inside each displayed interval, but instead on the event yields integrated in the corresponding signal regions together with the associated uncertainties. 

%############################################
%############################################
%############################################
\section{Statistical interpretation and expected sensitivity} \label{sec:stats}

In this section, we translate the selected signal and background event yields into an expected collider sensitivity in the $(\mxo,\mxt)$ mass plane. The interpretation is performed for the benchmark choice $c_B=\tilde c_B=1$ and $\Lambda=1~\mathrm{TeV}$ at $\sqrt{s}=13.6~\mathrm{TeV}$, using the selection and signal regions defined in \cref{subsec:selection}. Expected limits on the model are computed in terms of the signal-strength modifier,
\begin{equation}
  \mu = \frac{\sigma}{\sigma_{\rm th}},
\end{equation}
where $\sigma_{\rm th}$ denotes the nominal production cross section predicted by the EFT benchmark at a given point in the $(\mxo,\mxt)$ plane, including the radiative branching fraction $\mathrm{BR}(\xt\to\xo\gamma)$ and associated subleading offshell effects, and before the multiplication by the selection efficiency. The hypothesis $\mu=1$ therefore corresponds to the nominal signal prediction, while $\mu=0$ corresponds to the background-only case. Moreover, we consider the two statistical configurations introduced above, namely the inclusive single-SR counting region with $\met\ge200~\mathrm{GeV}$, and the three-SR configuration in which the same baseline-selected events are partitioned into the exclusive $\met$ intervals $\mathrm{SR}_1$, $\mathrm{SR}_2$ and $\mathrm{SR}_3$.

For each mass point, the statistical model is built from the event yields in the signal regions. In a bin $i$, the expected yield is written as
\begin{equation}
  \nu_i(\mu,\boldsymbol{\theta}) = \mu\,  s_i + b_i(\boldsymbol{\theta}),
\end{equation}
where $s_i$ and $b_i$ denote the nominal signal and background yields, and $\boldsymbol{\theta}$ collects the nuisance parameters encoding the background uncertainties. The likelihood is then constructed as a product of Poisson probabilities over the signal regions entering a given configuration,
\begin{equation}
  \mathcal{L}(\mu,\boldsymbol{\theta}) = \prod_i \mathrm{Pois}\!\left(n_i\,\middle|\,\nu_i(\mu,\boldsymbol{\theta})\right) \prod_k \pi_k(\theta_k),
\end{equation}
where $n_i$ denotes the event count in bin $i$, and where $\pi_k(\theta_k)$ denotes the constraint term associated with the nuisance parameter $\theta_k$. In this notation, the index $i$ runs over one bin in the single-SR configuration and over the three exclusive $\met$ bins in the three-SR configuration. For the inclusive single-SR configuration, we assign a $10\%$ uncertainty to the total SM background yield that we implement through a single Gaussian-constrained normalisation nuisance parameter. For the three-SR configuration, we assign one independent background-normalisation nuisance parameter to each $\met$ bin with relative uncertainties chosen as $\delta_b=(8.8\%,10.0\%,13.3\%)$ for the regions $\mathrm{SR}_1$, $\mathrm{SR}_2$ and $\mathrm{SR}_3$, respectively, guided by the total post-fit background uncertainties reported in~\cite{ATLAS:2025uij}. These uncertainties are applied to the total background yield in each bin and treated as uncorrelated between bins. We therefore do not model shape-dependent systematics that could induce migrations between the exclusive $\met$ regions and could be implemented through correlated nuisance parameters affecting the yields across the different signal regions.

We next use the profile-likelihood ratio
\begin{equation}
  q_\mu = -2\log  \frac{\mathcal{L}(\mu,\hat{\hat{\boldsymbol{\theta}}}_{\mu})}{\mathcal{L}(\hat{\mu},\hat{\boldsymbol{\theta}})},
\end{equation}
as a test statistic, where $\hat{\mu}$ and $\hat{\boldsymbol{\theta}}$ denote the unconditional maximum-likelihood estimators, while $\hat{\hat{\boldsymbol{\theta}}}_{\mu}$ denotes the conditional maximum-likelihood estimator of the nuisance parameters for a fixed value of $\mu$. Expected upper limits on $\mu$ at $95\%$ confidence level are derived with the modified frequentist $\mathrm{CL}_s$ prescription~\cite{Read:2002hq}, using an Asimov data set corresponding to the background-only hypothesis and evaluating the relevant $p$-values with the asymptotic formulae of~\cite{Cowan:2010js}. The reference luminosity used to normalise the background and signal event samples is taken to be $\mathcal{L}=139~\mathrm{fb}^{-1}$, matching the size of the full Run-2 ATLAS data set, while HL-LHC projections are obtained by rescaling the signal and background yields to $\mathcal{L}=3000~\mathrm{fb}^{-1}$. In both cases we keep the same object definitions, signal-region boundaries and relative systematic-uncertainty model so that this extrapolation should be understood as a controlled luminosity projection of the present prospective analysis rather than as a full HL-LHC optimisation including re-optimised selections.

\begin{figure}
  \centering
  \includegraphics[width=0.485\linewidth]{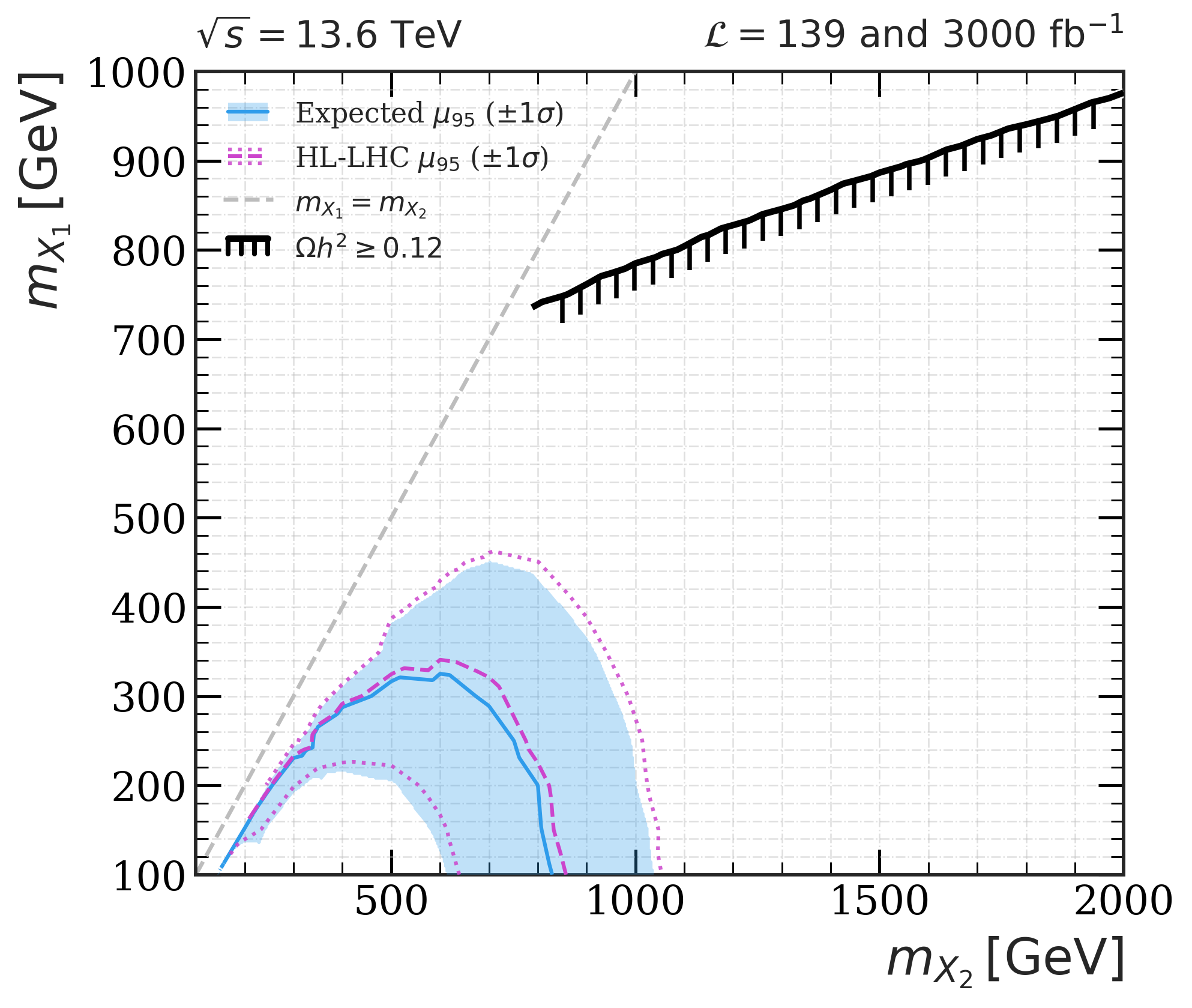}\hfill
  \includegraphics[width=0.485\linewidth]{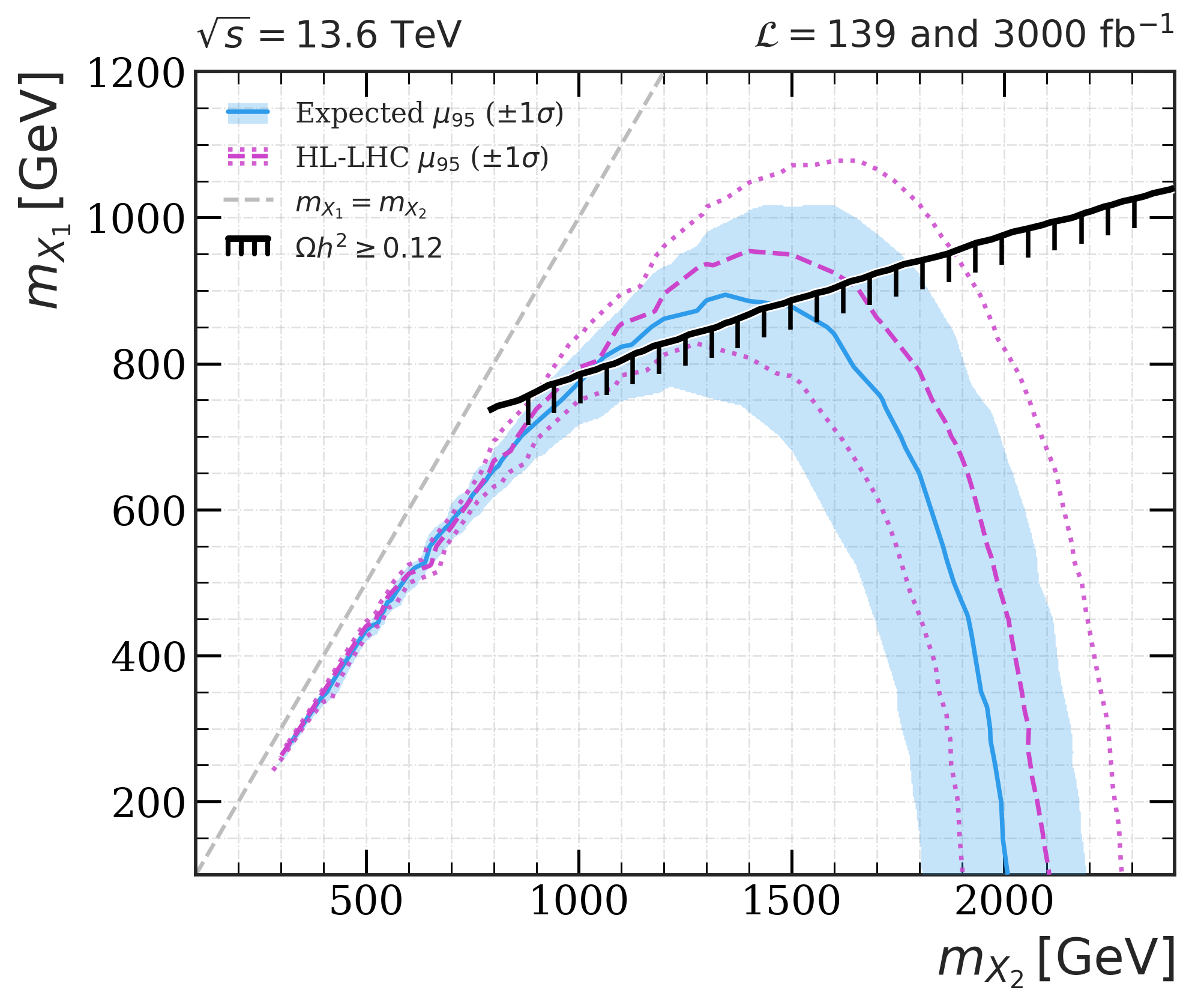}
  \caption{
    Expected $95\%$ CL upper limits on the signal-strength modifier $\mu$ in the $(\mxo,\mxt)$ plane, for the benchmark choice $c_B=\tilde c_B=1$ and $\Lambda=1~\mathrm{TeV}$ at $\sqrt{s}=13.6~\mathrm{TeV}$. We display the results obtained with the inclusive single-SR region (left) as well as those originating from the three-SR configuration (right). The solid blue and dashed magenta contours denote the median expected reaches defined by $\mu_{95}^{\rm exp}=1$ for $\mathcal{L}=139~\mathrm{fb}^{-1}$ and $3000~\mathrm{fb}^{-1}$, respectively, with the corresponding $\pm1\sigma$ expected bands shown by the blue shaded and magenta dotted regions. Points inside the contours where $\mu_{95}<1$ are expected to be excluded, and the black curve shows the freeze-out relic-density reference contour $\Omega_{\xo} h^2=0.12$.
  } \label{fig:limits}
\end{figure}

The resulting expected limits are shown in \cref{fig:limits}, in which the bands indicate the central $68\%$ expected variation of the $\mu_{95}=1$ exclusion contour under the background-only hypothesis given the uncertainty model described above. The comparison between the two panels demonstrates the importance of retaining the coarse $\met$-shape information through the three-SR configuration. In the inclusive single-SR analysis, the expected reach is limited to relatively light spectra, extending only to about $\mxo\simeq300~\mathrm{GeV}$ and $\mxt\simeq800-850~\mathrm{GeV}$ for $\mathcal{L}=139~\mathrm{fb}^{-1}$. This limited reach is a direct consequence of the inclusive counting strategy. The selected sample is dominated by the lower-$\met$ region, where the SM background rate is largest and where the harder signal tail is diluted by the total event count. Moreover, the single-SR exclusion region lies essentially entirely on the side of the freeze-out reference contour corresponding to an overabundant thermal relic. The single-SR analysis therefore only probes parameter points that would already overproduce DM in the standard thermal freeze-out mechanism, and hence be disfavoured if $\xo$ is required to account for the observed relic abundance. By contrast, the three-SR analysis improves the reach substantially by separating the low-$\met$ region which provides statistical power from the higher-$\met$ regions where the background is reduced and the signal-to-background ratio is enhanced for benchmarks with harder recoil spectra. As a result, the expected exclusion contour extends to significantly larger values of both $\mxo$ and $\mxt$, and for $\mathcal{L}=139~\mathrm{fb}^{-1}$, the three-SR configuration probes masses up to approximately $\mxo\simeq850-900~\mathrm{GeV}$ and $\mxt\simeq1.8-2.0~\mathrm{TeV}$. This reach intersects the freeze-out reference contour and extends into the region where the thermal relic abundance does not exceed the observed DM density. The three-SR analysis therefore probes benchmarks that are not already disfavoured by overclosure in the standard freeze-out scenario. 

The shape of the exclusion region follows from the physics discussed above. Close to the compressed regime, the photon and $\met$ spectra are too soft to pass the selection efficiently, whereas at very large $\mxt$ the gain in acceptance is overcome by the rapid fall of the production cross section. The maximal sensitivity is therefore obtained in an intermediate region where the mass splitting is large enough to produce hard visible objects and sizeable missing momentum, but the signal rate remains sizeable. The high-mass part of the three-SR reach should nevertheless be interpreted with the EFT-validity caveat discussed above. Since $\Lambda=1~\mathrm{TeV}$, mass points with TeV-scale dark vectors and hard selected events probe kinematic scales comparable to or larger than the EFT suppression scale. As already mentioned, the corresponding contours should therefore be viewed as benchmark sensitivities to the EFT framework, while a fully conservative interpretation would require either an event-by-event truncation in the relevant scale or an explicit UV completion.

The HL-LHC projection extends the reach further. The improvement is modest in the single-SR configuration, while the three-SR reach extends to about $\mxo\simeq950~\mathrm{GeV}$ and $\mxt\simeq2.2~\mathrm{TeV}$. The improvement is nevertheless not uniform across the mass plane. In regions where the analysis is statistically limited, the larger luminosity directly increases the sensitivity, while in regions dominated by the uncertainties or by the steeply falling signal rate the gain is more moderate. 

Finally, the overlay of the freeze-out contour should be interpreted as a cosmological reference for the same benchmark parameters, rather than as an additional constraint. Points on the contour reproduce the observed relic abundance in the standard thermal freeze-out scenario, while points away from it may correspond to underabundant or overabundant thermal histories. Moreover, no freeze-in contour is overlaid, because in that scenario the relic abundance fixes a relation involving $\Lambda$ and $T_{\rm rh}$ rather than a unique curve in the $(\mxo,\mxt)$ plane for the fixed collider benchmark $\Lambda=1~\mathrm{TeV}$. Since the collider rate scales approximately as $\Lambda^{-4}$, freeze-in benchmarks requiring multi-TeV values of $\Lambda$ are correspondingly more difficult to probe, whereas low-$T_{\rm rh}$ scenarios requiring $\Lambda$ closer to the TeV scale are the most relevant freeze-in targets for collider searches.

%############################################
%############################################
%############################################
\section{Conclusion \label{sec:conclusions}}

In this work, we have investigated the phenomenology of a two-vector dark-sector effective theory in which the lightest state $\xo$ is stabilised by a dark parity and constitutes the DM candidate, while the heavier state $\xt$ communicates with the SM through dimension-six operators involving the hypercharge field strength. After electroweak symmetry breaking, these interactions induce the radiative decay $\xt\to\xo\gamma$ and, when kinematically allowed, the competing mode $\xt\to\xo Z$. In the prompt-decay regime considered in this study, the associated production of the two dark vectors in the presence of QCD radiation therefore leads to a characteristic $\gamma+\mathrm{jets}+\met$ LHC signature.

We first analysed the decay properties of the heavier dark vector and the cosmological implications of the model. For the considered benchmark choice, the $\xt$ lifetime remains far below detector scales throughout the region relevant for the collider study, justifying a prompt-photon treatment. The radiative branching fraction is equal to unity below the on-shell $Z$ threshold and remains sizeable above threshold, approaching the pattern expected from the electroweak decomposition of the hypercharge field strength at large mass splittings. On the cosmology side, the same effective interactions control the freeze-out and freeze-in production of the DM candidate. In the freeze-out scenario, the observed relic density selects a reference contour in the $(\mxo,\mxt)$ plane for fixed EFT parameters, while in the freeze-in scenario the relic abundance also depends on the reheating temperature and therefore does not define a unique contour for a fixed collider benchmark.

We then performed a prospective collider analysis at $\sqrt{s}=13.6~\mathrm{TeV}$, using Monte Carlo simulations interfaced with parton showering, hadronisation and a fast detector simulation. The event selection targets a prompt isolated photon, at least one recoil jet and sizeable missing transverse momentum, together with lepton and $b$-jet vetoes to suppress the SM background. We compared an inclusive single-SR counting analysis and a three-SR analysis based on exclusive $\met$ intervals which retains coarse information on the hardness of the event, separating the statistically powerful low-$\met$ region from higher-$\met$ regions where the SM background is reduced. The inclusive single-SR strategy reaches only relatively light spectra for $\mathcal{L}=139~\mathrm{fb}^{-1}$ and $3000~\mathrm{fb}^{-1}$, and this reach lies on the overabundant side of the freeze-out reference contour. Therefore, within the standard thermal freeze-out interpretation, it probes only parameter points that would already overproduce dark matter if $\xo$ were required to account for the observed relic abundance. By contrast, the three-SR strategy extends the reach to significantly larger masses, and the expected exclusion intersects the freeze-out reference contour and extends into the region where the thermal relic abundance does not exceed the observed density, therefore probing benchmark points that are not already disfavoured by overclosure in the standard freeze-out scenario.

Overall, our results show that photon-plus-jets-plus-missing-momentum searches provide a sensitive probe of dimension-six portals between the SM and vector dark sectors. The analysis highlights in particular the importance of exploiting even coarse differential information in the missing transverse momentum. Whereas an inclusive strategy misses the cosmologically interesting freeze-out region, a simple three-bin treatment can reach parameter space compatible with the observed relic abundance, and further optimisation of the $\met$ binning could possibly sharpen this sensitivity. Our results should however be interpreted within the limitations of the EFT approach, as the high-mass part of the reach probes in particular dark-vector masses and selected-event kinematics that can be comparable to or larger than the benchmark EFT scale.

%############################################
%############################################
%############################################
\begin{acknowledgments}
B.F., M.D.G., B.L. J.O. and Y.V. were supported in part by Grant ANR-21-CE31-0013, Project DMwithLLPatLHC, from the French \textit{Agence Nationale de la Recherche} (ANR) and Simons Foundation (Award Number:1023171-RC). B.L. was also partly supported by a mobility grant from the Brazilian \textit{Conselho Nacional de Desenvolvimento Científico e Tecnológico} (CNPq). FSQ was supported by Simons Foundation (Award Number:1023171-RC), FAPESP Grant 2018/25225- 9, 2021/01089-1, 2023/01197-4, ICTP-SAIFR FAPESP Grants 2021/14335-0, CNPq Grants 307130/2021-5, and ANID-Millennium Science Initiative Program ICN2019\_044, and FINEP under the project 213/2024. 
\end{acknowledgments}
\bibliographystyle{JHEP}
\bibliography{ref}
%############################################
%############################################
%############################################
\end{document}